\documentclass[12pt]{iopart}
\usepackage{iopams}
\usepackage{epsfig}
\expandafter\let\csname equation*\endcsname\relax
\expandafter\let\csname endequation*\endcsname\relax
\usepackage{amsmath}
\usepackage{amssymb}
\usepackage{mathtools}
\usepackage{color}
\usepackage{ulem}

\begin{document}

\title[A reversed form of public goods game]{A reversed form of public goods game: equivalence and difference}

\author{Chaoqian Wang$^{1}$ and Attila Szolnoki$^{2,*}$}

\address{$1$ Department of Computational and Data Sciences, George Mason University, Fairfax, VA 22030, USA\\
$2$ Institute of Technical Physics and Materials Science, Centre for Energy Research, P.O. Box 49, H-1525 Budapest, Hungary\\
$*$ Corresponding author\\}

\ead{CqWang814921147@outlook.com; szolnoki.attila@energia.mta.hu}

\begin{abstract}
According to the public goods game (PGG) protocol, participants decide freely whether they want to contribute to a common pool or not, but the resulting benefit is distributed equally. A conceptually similar dilemma situation may emerge when participants consider if they claim a common resource but the related cost is covered equally by all group members. The latter establishes a reversed form of the original public goods game (R-PGG). In this work, we show that R-PGG is equivalent to PGG in several circumstances, starting from the traditional analysis, via the evolutionary approach in unstructured populations, to Monte Carlo simulations in structured populations. However, there are also cases when the behavior of R-PGG could be surprisingly different from the outcome of PGG. When the key parameters are heterogeneous, for instance, the results of PGG and R-PGG could be diverse even if we apply the same amplitudes of heterogeneity. We find that the heterogeneity in R-PGG generally impedes cooperation, while the opposite is observed for PGG. These diverse system reactions can be understood if we follow how payoff functions change when introducing heterogeneity in the parameter space. This analysis also reveals the distinct roles of cooperator and defector strategies in the mentioned games. Our observations may hopefully stimulate further research to check the potential differences between PGG and R-PGG due to the alternative complexity of conditions.
\end{abstract}



\maketitle

\section{Introduction}
\label{introduction}

When an individual has to pay for the consumption of goods or services then there is a natural incentive to moderate this consumption. But there are cases when the relationship between the consumption and the emerging cost is more subtle because both actions are made by a group of persons. For example, in a university dormitory or a shared household each member can use electricity or water freely, but all members share the related bills equally. This collective responsibility may induce less trustworthy individual consumption because members may easily consume more than they would if the personal consumption is recorded and the related bill is paid separately. As a result, they can easily become careless because the extra use is paid collectively. Notably, the reverse is also true: an individual saving of usage would result in just a modest decrease in an individual bill because others will also benefit from a responsible act. On the other side, if all participants behave trustworthily and consume just the necessary resource then they would pay the minimal cost. In parallel, they could still enjoy the benefit of a collective venture, like maintaining a less expensive joint infrastructure. This establishes a dilemma between individual and collective interests when players are motivated to defect and consume more than they would individually. A similar dilemma occurs at a larger scale when countries use the same natural resources, like the oceans or the atmosphere~\cite{milinski_pnas08,hilbe_prsb10,pacheco_plrev14,sun_ww_is21}.

The conflict of individual and collective interests, often called as a social dilemma, is the central problem of several game-theoretical models, including prisoner's dilemma game \cite{szabo1998evolutionary,takeshue_epl19,amaral_pre21,zhu_pc_epjb21}, snowdrift game \cite{hauert2004spatial,chen_xj_epl10,graeser_njp11}, stag-hunt game \cite{skyrms2004stag,starnini_jsm11,deng_ys_epjb22}, ultimatum game \cite{page2000spatial,sinatra_jstat09,szolnoki_prl12,chen_w_pa19}, trust game \cite{berg1995trust,chica_srep19,zheng_lp_pa21}, donation game \cite{nowak2006evolutionary,li_xy_epjb20,yang_hx_pa20}, or recently proposed involution game \cite{wang2021replicator,wang2022modeling,wang_cq_amc22}, including the $N$-person versions of these games \cite{zheng2007cooperative,souza2009evolution,li_k_csf21,pacheco2009evolutionary,luo2021evolutionary,chen_w_srep17,chica_cnsns19}. The common feature of all related situations, which are captured by the mentioned mathematical games, is that cooperation would provide the highest collective benefit, but players can gain more individually if they decide to defect and exploit others \cite{sigmund2010calculus,hilbe_n18,couto_njp22}. However, from the viewpoint of our present work, the most important version is the so-called public goods game (PGG) played by several players simultaneously \cite{perc2013evolutionary,alvarez2021evolutionary,lv_amc22,kang_hw_pla21,shen_y_pla22,liu_lj_rspa22}. In this framework, group members decide independently whether they want to contribute to a common pool or not. The accumulated contributions are then enhanced by an $r>1$ productivity factor which expresses the synergy among collaborators: their collective efforts are more effective than a simple sum of their contributions. Finally, the enhanced sum is distributed among all group members equally regardless of whether they contributed to the common pool earlier. It is worth noting that the original public good dilemma can be framed from different aspects: we may establish or maintain a common pool where both goals generate the same individual dilemma \cite{gaechter17}.

In the original PGG, the contribution, which involves direct negative income to a participant, is optional, while the benefit is universal for all group members. From this viewpoint, the situation we described earlier is the opposite: consumption, which means a positive personal income, is optional while the contribution to the resulting cost is mandatory. Put differently, in a PGG the negative part is optional and the positive part is universal. In contrast, in our present case the positive part is optional and the negative part is universal. This is why we can call this setup a {\it reversed} public goods game (R-PGG) where cooperation means to consume the minimal from the common resource, while a defector chooses to consume the pool extravagantly and wastefully. 

In this work, we explore the proposed R-PGG model and reveal its relation to the original PGG. We therefore study the equilibrium and evolutionary behavior of competing strategies and investigate how it is possible to reduce the undesired overuse of common resources. Evidently, this is the core question of evolutionary game theory which targets to understand the evolution of cooperation among self-interested competitors \cite{nowak2004emergence,quan_j_pa21,su2018understanding,fu_mj_pa21,ohdaira_srep22}.
Besides a well-mixed population, studied by replicator dynamics \cite{schuster1983replicator,duong_dga20,wang_q_amc18,liang_rh_pre22}, we also consider structured populations, which could be a decisive factor in how evolution proceeds \cite{nowak1992evolutionary,ohtsuki2006simple,perc2017statistical,allen2017evolutionary,flores_jtb21,quan_j_c19,wei_x_epjb21}. Notably, we not only check the consequences when players are arranged on a lattice topology, but more realistic topologies, including small-world and scale-free features are also considered.

Moreover, we also assume more complex conditions, including the heterogeneity of key parameters, and explore their impacts on the evolutionary outcome. Indeed, heterogeneity in various forms is not simply a more realistic approach, but could also be a decisive factor in how cooperation evolves \cite{wang_cq_amc22,perc_pre08,santos_n08}. For the public goods game, the heterogeneity of allocation \cite{lei2010heterogeneity,szolnoki_amc20,lee_hw_pa21}, productivity \cite{liu_jz_epjb21,rong_zh_c19,deng_ys_pa21}, input \cite{yuan2014role,huang2015effect,weng2021heterogeneous,ma2021effect}, and their combinations have been studied intensively \cite{zhang2017impact,fan2017promotion,liu2021effects}. In particular, Hauser {\it et al.} \cite{hauser2019social} systematically analyzed public goods games among unequals by equilibrium calculation, evolutionary analysis, and human experiments, where heterogeneous game parameters depict unequal individuals. While heterogeneity can sometimes be a cooperator-supporting condition, it could also lead to inequality \cite{mcavoy2020social}.

The remaining of our paper is organized as follows. In the next section we define the R-PGG model and discuss its extension to heterogeneous situations. Section~\ref{equivalence} compares the original and reverse homogeneous models for well-mixed and spatial structured populations where their equivalence is demonstrated. In Sec.~\ref{difference} we present the heterogeneous case where the surprising difference between the models is revealed. We here also present the mechanisms which explain the equivalence and difference between their behaviors. In the last section we discuss our findings and provide some potential applications to alternative real-life situations. Our arguments are greatly supported by an intensive Appendix where all complementary observations are presented.

\section{Model}
\label{model}

We here define the R-PGG model in analogy with the original PGG. First, we consider a specific case, where players are homogeneous (Sec.~\ref{homomodel}), then we extend our model to a more general case, where the players are not necessarily equal (Sec.~\ref{hetemodel}).

\subsection{The underlying model}
\label{homomodel}

Similarly to PGG, in an R-PGG a group is formed by $g$ members. A player is free to require goods, but all involved members should cover the emerging cost of the goods equally. For simplicity, we assume binary strategies: (i) not requiring goods (cooperation, $C$); (ii) requiring goods (defection, $D$). Later we explain why these strategies can be considered as cooperation or defection. If a player defects and requires goods, she receives goods valued $m$ ($m>0$). Otherwise, she receives nothing. When all group members made their independent decisions about their consumptions, each player equally pays a cost to cover the goods required by the whole group. The payoff $\pi_C$ and $\pi_D$ for a cooperative or for a defective player can be calculated as 
\begin{subequations}\label{payoffCD}
	\begin{align}
		\pi_C&=-\frac{g_D m\lambda}{g} \label{payoffC}\\
		\pi_D&=m-\frac{(g_D+1) m\lambda}{g} \label{payoffD}\,,
	\end{align}
\end{subequations}
where there are $g_D$ other defectors among the neighbors of the focal player. Here $\lambda > 1$ parameter denotes the waste factor associated with requiring public goods. The larger the waste factor, the greater the loss caused by the consumption. If  $\lambda<g$ then requiring goods always brings positive income to a defective individual. Therefore, to meet a proper dilemma, we assume that $1<\lambda<g$. 

\subsection{The extension to heterogeneous players}
\label{hetemodel}

More generally, we can consider heterogeneous $m$ and $\lambda$ game parameters for different players. A player $i$ ($i=1,2,\dots,g$) in a group may require goods valued $m_i$ ($m_i>0$), with its waste factor $\lambda _i$. In the extended version, we introduce a continuous strategy space, hence the strategy of player $i$ can be denoted by a proportion $y_i$ ($0\leq y_i\leq 1$) of goods $m_i$, where $y_i=0$ means full cooperation and $y_i=1$ is full defection. If not mentioned otherwise, we use the pure strategy set, $y_i \in \{ 0,1\}$, in this work. The payoff $\pi_i$ of player $i$ can be calculated as
\begin{equation} \label{unequalpayoff}
	\pi_i=y_i m_i-\frac{1}{g}\sum_{j=1}^g y_j m_j\lambda_j\,,
\end{equation}
where $j$ goes through every player in the group, including player $i$. 

In this extension, $m_i$ measures the full consumption capacity of player $i$. In real-world scenarios, an individual's needs and capacity may vary. Secondly, $\lambda_i$ characterizes the ``effectiveness" of player $i$ in spoiling the required public goods. Note that players with the same consumption capacity and strategy may generate different final cost values due to individual efficiency.

\section{Equivalence between R-PGG \& PGG}
\label{equivalence}

We first demonstrate that the proposed R-PGG model is equivalent to the PGG in many aspects. The static analysis shows the equivalence in general homogeneous cases. From the perspective of evolutionary dynamics, the equivalence of the underlying homogeneous model also holds both in well-mixed and structured populations.

\subsection{Static analysis among unequals}
\label{unequals}

Hauser {\it et al.}~\cite{hauser2019social} proposed a general framework for the PGG among unequal participants (i.e., players with heterogeneous game parameters). We here show that our R-PGG model fits this general framework. In the following, we perform an equilibrium analysis.

According to the general heterogeneous case defined by Eq.~(\ref{unequalpayoff}), we denote the requirement vector by $\mathbf{y}=(y_1, y_2, ..., y_g)$ in a group of $g$ players, and assume that $\mathbf{y}>\mathbf{y}'$ if $y_i>y'_i$ for all $i \in \{1,\dots g\}$ group members. We also denote the full consumption or desire profile by vector $\mathbf{m}=(m_1, m_2,\dots, m_g)$ and the waste vector $\boldsymbol{\lambda}=(\lambda_1, \lambda_2,\dots, \lambda_g)$ for the actual group. These $\mathbf{m}$, $\boldsymbol{\lambda}$, and $\mathbf{y}$ vectors determine the payoff values $\pi(\mathbf{m},\boldsymbol{\lambda},\mathbf{y})\in \mathbb R^g$ 
for all group members. 
In addition, we denote the accumulated group desire or maximal group consumption by $M\coloneqq \sum_{i=1}^g m_i$, and the total payoff of all involved players by $U(\mathbf{m},\boldsymbol{\lambda},\mathbf{y})\coloneqq \sum_{i=1}^g \pi_i(\mathbf{m},\boldsymbol{\lambda},\mathbf{y})$.

In Ref.~\cite{hauser2019social}, the authors introduced a so-called ``cooperation vector", denoted by $\mathbf{x}=(x_1, x_2,\dots, x_g)$, which characterizes the cooperation profile of all involved participants. In our R-PGG model, consumption is the focus, therefore it is appropriate to use a ``defection vector" which practically determines the individual aims to consume their maximal capacities via $\mathbf{y}=(y_1, y_2, \dots, y_g)$ requirement vector. Since cooperation and defection complement each other, the sum of these vectors is always constant and gives a unit value for every $i$ component. Namely, $\mathbf{y}+\mathbf{x}=\mathbf{1}$.
In strong analogy with the heterogeneous PGG model~\cite{hauser2019social}, the following properties are valid here, which describe the conflict between the collective and individual incentives on how to utilize public resources:
\begin{itemize}
	\item
	\textbf{Continuity:} 
	The payoff function $\pi(\mathbf{m},\boldsymbol{\lambda},\mathbf{y})$ is continuous in both the arguments $\mathbf{m}$ and $\mathbf{y}$.
	\item 
	\textbf{Negative externalities:}
	If $\mathbf{y}$ and $\mathbf{y}'$ are two requirement vectors such that $y_i=y'_i$ and $y_j\geq y'_j$ for $\forall j\neq i$, then $\pi_i(\mathbf{m},\boldsymbol{\lambda},\mathbf{y})\leq \pi_i(\mathbf{m},\boldsymbol{\lambda},\mathbf{y'})$ for $\forall \mathbf{m}$. The inequality $\pi_i(\mathbf{m},\boldsymbol{\lambda},\mathbf{y})< \pi_i(\mathbf{m},\boldsymbol{\lambda},\mathbf{y'})$ holds iff $\exists j$ such that $m_j>0, y_j>y'_j$.
	\item 
	\textbf{Incentives to free-require:} 
	If $\mathbf{y}$ and $\mathbf{y}'$ are two requirement vectors such that $y_i>y'_i$ and $y_j=y'_j$ for $\forall j\neq i$, then $\pi_i(\mathbf{m},\boldsymbol{\lambda},\mathbf{y})\geq \pi_i(\mathbf{m},\boldsymbol{\lambda},\mathbf{y'})$ for $\forall \mathbf{m}$. The inequality $\pi_i(\mathbf{m},\boldsymbol{\lambda},\mathbf{y})> \pi_i(\mathbf{m},\boldsymbol{\lambda},\mathbf{y'})$ holds iff the required goods are positive, $m_i>0$.
	\item 
	\textbf{Non-optimality of defection:}
	If $\mathbf{y}$ and $\mathbf{y}'$ are two requiring vectors such that $\mathbf{y}\leq \mathbf{y}'$, then $U(\mathbf{m},\boldsymbol{\lambda},\mathbf{y})\geq U(\mathbf{m},\boldsymbol{\lambda},\mathbf{y'})$. The inequality $U(\mathbf{m},\boldsymbol{\lambda},\mathbf{y})> U(\mathbf{m},\boldsymbol{\lambda},\mathbf{y'})$ holds iff $\exists i$ such that $m_i>0, y_{i}<y'_i$.	
\end{itemize}

The so-called Grim player $i$ cooperates in the initial round ($y_i=0$). Then, if all players cooperate as well, player $i$ keeps cooperating. Otherwise, if any player defects in previous rounds, player $i$ defects and keep defecting in the future ($y_i=1$) \cite{sigmund2010calculus}. In anology to Ref.~\cite{hauser2019social}, when all players are Grim players, it is a subgame perfect equilibrium for the given requirement distribution, which means that full cooperation can be sustained among Grim players. Given that all players follow Grim strategy, the following condition must hold for all players $i$ to sustain cooperation:
\begin{equation}\label{grimcondi}
	\delta(\pi_i(\mathbf{m},\boldsymbol{\lambda},\mathbf{0}_{+i})-\pi_i(\mathbf{m},\boldsymbol{\lambda},\mathbf{1}))\geq \pi_i(\mathbf{m},\boldsymbol{\lambda},\mathbf{0}_{+i})-\pi_i(\mathbf{m},\boldsymbol{\lambda},\mathbf{0})\,,
\end{equation}
where $\delta$ is the probability of another round. Here vector $\mathbf{1}$ means all players defect, $\mathbf{0}$ means all players cooperate, and $\mathbf{0}_{+i}$ means that all players cooperate but player $i$ defects. According to Eq.~(\ref{grimcondi}), the expected benefit from the future cooperation of others must exceed the incentive to defect in the current round.

Based on Eq.~(\ref{unequalpayoff}), the individual payoff value is
\begin{eqnarray}\label{rpgggrim}
	\pi_i(\mathbf{m},\boldsymbol{\lambda},\mathbf{y})&=-\frac{1}{g}\sum_{j} y_j m_j \lambda_j+m_i y_i \nonumber\\
	&=-\frac{1}{g}\sum_{j\neq i}y_j m_j \lambda_j +m_i(1-\frac{\lambda_i}{g})y_i \,.
\end{eqnarray}
Therefore, 
\begin{align}\label{rpgguvalue}
\begin{cases} 
	\displaystyle{\pi(\mathbf{m},\boldsymbol{\lambda},\mathbf{1})\,\,\,\,\,=-\frac{1}{g}\sum_{j\neq i}m_j\lambda_j +m_i(1-\frac{\lambda_i}{g})}\\
	\displaystyle{\pi(\mathbf{m},\boldsymbol{\lambda},\mathbf{0})\,\,\,\,\,=0}\\
	\displaystyle{\pi(\mathbf{m},\boldsymbol{\lambda},\mathbf{0}_{+i})=m_i(1-\frac{\lambda_i}{g})}\,.
\end{cases}
\end{align}
By substituting Eqs.~(\ref{rpgguvalue}) into Eq.~(\ref{grimcondi}), the necessary and sufficient condition for the feasibility of cooperation for all involved player $i$ is
\begin{equation}\label{rpgggrimcondi}
	\frac{\delta}{g}\sum_{j\neq i}m_j \lambda_j \geq m_i (1-\frac{\lambda_i}{g})\,.
\end{equation}

Like the PGG, repeated games (indicated by a larger $\delta$) promote cooperation. For example, according to Eq.~(\ref{rpgggrimcondi}), given equal desire $\mathbf{m}=(M/g,\dots,M/g)$, full cooperation is feasible iff
\begin{equation}\label{rpggdeltacondi}
	\delta\geq \frac{g-\min{(\lambda_j)}}{\sum_{j=1}^{g}\lambda_j-\min{(\lambda_j})}\,.
\end{equation}

We can deduce the optimal requirement vector $\mathbf{m}$, maximally cooperative when requiring the lowest continuation probability $\delta$ for cooperation to be feasible. In particular, if we consider only two players, then full cooperation is feasible iff  
\begin{equation}\label{rpggtwoplayercondi}
	\delta\geq \max{\left(\frac{m_1(2-\lambda_1)}{m_2\lambda_2},\frac{m_2(2-\lambda_2)}{m_1\lambda_1}\right)}\,.
\end{equation}
To minimize the right-hand side of Eq.~(\ref{rpggtwoplayercondi}), we have the condition
\begin{equation}\label{rpggmaxcondi}
	\frac{m_1}{m_2}=\sqrt{\frac{\lambda_2(2-\lambda_2)}{\lambda_1(2-\lambda_1)}}\,,
\end{equation}
which maximizes cooperation.

We can see that in the conditions of Eq.~(\ref{rpgggrimcondi}), Eq.~(\ref{rpggdeltacondi}), and Eq.~(\ref{rpggmaxcondi}), the parameter $\lambda$ plays the same role as $r$ in PGG, and parameter $m$ plays the same role as $c$ in PGG (see Ref.~\cite{hauser2019social} for the corresponding conditions in PGG and note that they denote the contribution $c$ by endowment $e$). In other words, R-PGG is equivalent to PGG in the framework of social dilemmas among unequals.

\subsection{Evolutionary dynamics in a well-mixed population}
\label{unstruc}

The evolutionary game theory focuses on how cooperation can evolve via a dynamic process. Unlike the traditional game theory, discussed in Sec.~\ref{unequals}, here we adopt the simplest strategy set (a player employs unconditional cooperation or defection) \cite{nowak1992evolutionary}.
In particular, according to microscopic dynamics, a strategy with a higher payoff value can be reproduced with a higher probability via a certain dynamical rule, including, but not limited to death-birth \cite{ohtsuki2006simple}, birth-death \cite{ohtsuki2006simple}, pairwise comparison \cite{szabo1998evolutionary}, and imitation \cite{ohtsuki2006simple} rules. For simplicity, we here apply the pairwise comparison dynamics with Fermi-function-type probability. Accordingly, player $i$ with payoff $\pi_i$ adopts the strategy of player $i'$ who has payoff $\pi_{i'}$ with the probability
\begin{equation}\label{fermi}
	Q(\pi_i \gets \pi_{i'})=\dfrac{1}{1+\e^{-\omega(\pi_{i'}-\pi_i)}}\,.
\end{equation}
Here, $\omega$ ($\omega>0$) is considered as the strength of selection. A stronger selection strength $\omega$ means that players are more sensitive to the payoff difference. The higher the payoff of player ${i'}$, the higher probability is that player $i$ imitates the strategy of player ${i'}$.

Without loss of generality, we first study the evolutionary dynamics of R-PGG in a well-mixed population of homogeneous players. 
In Appendix~A1 we first apply deterministic dynamics and find two solutions to the replicator equation. They are full defector and full cooperator states. Their stability depends on the sign of the payoff difference between the competing strategies. To compare the results with the PGG case, we obtain equivalent expressions if we replace the original $r$ and $c$ parameters with $\lambda$ and $m$. Hence the homogeneous R-PGG is equivalent to homogeneous PGG in the framework of replicator dynamics in an infinite and unstructured population.

We can also consider stochastic dynamics in a finite population of size $N$, as specified in Appendix~A2. By calculating the expected payoff values for both strategies, the transition matrix of the Markov process can be determined. Similarly to the above-discussed approach, the process has two absorbing states. In the $\omega \to 0$ weak selection limit, the fixation probability agrees with the one obtained for the original PGG if we replace $\lambda$ and $m$ parameters with $r$ and $c$. Therefore, we can conclude that homogeneous R-PGG \& PGG are also equivalent in the framework of stochastic dynamics in a finite and unstructured population.

\subsection{Evolutionary dynamics in structured populations (the homogeneous case)}
\label{struc}

Now, we consider R-PGG on a graph $\mathcal{G}=(\mathcal{N},\mathcal{L})$ of a finite and structured population $N$. The graph $\mathcal{G}$ is unweighted and undirected. $\mathcal{N}$ is the node set of the graph, $\mathcal{N}=\{1,2,\dots,N\}$ and each node represents a player. $\mathcal{L}$ is the link set of the graph, $\mathcal{L}=\{l_1,l_2,\dots,l_{|\mathcal{L}|}\}$. We denote by $l_{i_1 i_2}$ a link with order 2, connecting two nodes $i_1$ and $i_2$. 
The size of $\mathcal{L}$ varies with specific graphs (e.g., for a complete graph, the size of $\mathcal{L}$ is $|\mathcal{L}|=N(N-1)/2$). The degree of node $i$ (i.e., the number of link(s) containing $i$) is denoted by $k_i$. Self-loop is not allowed.

At each elementary Monte Carlo (MC) step, a random node $i\in \mathcal{N}$ is selected. If $k_i=0$, nothing happens. If $k_i\neq 0$, then a random link $l_{ii'}$ is selected between the focal player $i$ and a neighbor $i'$. We then calculate their payoff values. Player $i$ participates in $k_i+1$ R-PGGs centered on herself and all her neighbor(s). First, the contribution is calculated from the $g_i=k_i+1$-size group centered on player $i$. After, we calculate the income from each group centered on an $i$'s neighbor(s). Last, player $i$ accumulates these  payoff values to reach the total sum
\begin{align}\label{graphpayoff}
	\pi_i=&~\frac{1}{k_i+1}\left[ y_im_i-\frac{1}{g_i}\left(y_im_i\lambda_i+\sum_{l_{ij_1}\in\mathcal{L}}y_{j_1}m_{j_1}\lambda_{j_1}\right)\right.\nonumber
	\\
	&\left.+\sum_{l_{ij_1}\in \mathcal{L}}\left(y_i m_i-\frac{1}{g_{j_1}}\left(y_im_i\lambda_i+\sum_{l_{j_1j_2}\in \mathcal{L}} y_{j_2} m_{j_2}\lambda_{j_2}\right)\right)\right]\,.
\end{align}

The $\pi_{i'}$  payoff of player $i'$ can be calculated in a similar way. Then, player $i$ adopts the strategy of model player $i'$ with the probability $Q(\pi_{i}\gets\pi_{i'})$ defined by Eq.~\ref{fermi}. Each full time step contains $N$ elementary MC steps specified above.

In the weak selection limit, we can determine the $\lambda^\star$ threshold value for the success of cooperation through the identity-by-descent method introduced by Allen and Nowak \cite{allen2014games}. The details can be found in 
Appendix~B. This calculation suggests that the threshold $\lambda^\star$ in R-PGG is exactly the same as $r^\star$ for the original PGG (see Refs.~\cite{su2018understanding,su2019spatial} for the deduction of $r^\star$ in the original PGG). In other words, homogeneous R-PGG \& PGG predict equivalent threshold values for the emergence of cooperation in the weak selection limit.

For non-transitive graphs, which may lead to varying $k_i$ degree among different players $i$, the explicit threshold of cooperation emergence in multiplayer games remains unsolved (only the results in two-player games have been solved \cite{allen2017evolutionary}). In this case, we may use the average degree $\langle k\rangle=(\sum_{i=1}^N k_i)/N$ to estimate the threshold value. In this case, the average number of players in each group is $\langle g\rangle =\langle k\rangle+1$.

\begin{itemize}
	\item 
	\textbf{Monte Carlo simulations}
\end{itemize}

To extend our study, we also apply numerical simulations which can be done for arbitrary selection strength. Now we choose $\omega=10$ (i.e., $1/\omega=0.1$) and $N=40000$. This intermediate level of selection strength was frequently applied by previous works \cite{lv_amc22,kang_hw_pla21,shen_y_pla22,deng_ys_pa21,szolnoki_pre09c,gracia-lazaro_pre14,yang_hx_pa19,meloni_rsos17,szolnoki_epl10}.
At the beginning, we randomly assign each agent's strategy by cooperation or defection. Then, we simulate for $15000$ MC steps, and measure the  $\langle x \rangle$ stationary portion of cooperators. This can be done by averaging $x_i$ values over the last $5000$ steps of the simulations.

To cover all significantly different interaction topologies, we apply both regular and heterogeneous graphs. In the former case we use an $L \times L$ square lattice with periodic boundary conditions where players have four nearest neighbors, hence they form 5-member groups. The typical linear system size was $L=200$. Furthermore, we also use  Watts-Strogatz-type (WS) small-world topology where the graph is generated from a ring of $N$ nodes having $k=4$ bonds for each player \cite{watts1998collective}. Then, each link is rewired randomly with probability $p=0.5$, where self-loop and double links are forbidden. Finally, we also use Barab{\'a}si-Albert-type (BA) scale-free graphs where the graph is generated starting from a 3-node complete graph and new nodes are attached to the network with two new links \cite{barabasi1999emergence}. In this way we keep $\langle k\rangle=4$ average degree, which makes the results comparable for all cases.

\begin{figure}[h!]
	\centering
	\makebox[\textwidth][c]{\includegraphics[width=14.5cm]{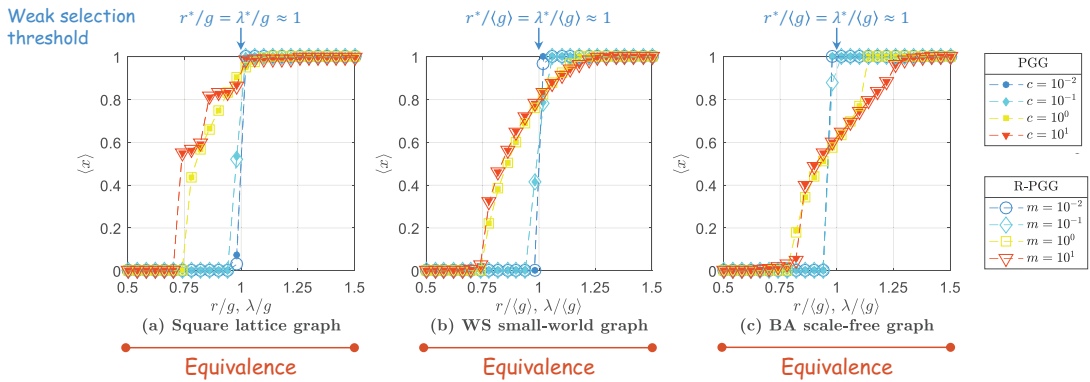}}\\
	\caption{Cooperation level as a function of normalized synergy factor $r/\langle g\rangle$ and waste factor $\lambda/\langle g\rangle$. Panels show the results on different graphs, as indicated in the titles. In addition, we mark by arrows the threshold level of the control parameter favoring cooperation under weak selection.}\label{graph r}
\end{figure}

Our observations are summarized in Fig.~\ref{graph r} where we present the $\langle x\rangle$ cooperation level both for PGG and R-PGG models. Evidently, as we pointed out earlier, parameters $r$ and $\lambda$ serve as control parameters in the former and latter cases, respectively. The overlap between the results of the models is convincing. Evidently, there are differences between the applied topologies, but both PGG and R-PGG change identically. A similar conclusion can be drawn about the influence of cost $c$ and consumption $m$ parameters. If one can require more goods (by increasing $m$), then the cooperation proportion is more sensitive to the waste factor (i.e., increasing with $\lambda$ more gradually). On the one hand, this leads to the emergence of cooperation, $\langle x\rangle>0$, at a smaller $\lambda$ value. On the other hand, it also leads to the dominance of cooperation, $\langle x\rangle=1$, at a greater $\lambda$. The latter effect is more evident on the small-world graph than on the square lattice graph, and is most evident on the scale-free graph.

According to Eq.~(\ref{graphpayoff}), $m$ can be extracted as a common factor in every $\pi_i$ for the homogeneous case. Then, $m$ plays the same role as $\omega$ in the Fermi-function defined by Eq.~(\ref{fermi}). In other words, the selection strength is also directly measured by $m$. The same inference holds in PGG, where the selection strength can be measured by $c$. Taking $c=m=10^{-2}$, we have the results valid under a weak selection in numerical simulations. Using $N=40000$ and $k=4$ in Eq.~(\ref{emergelimit}), we have $\lambda^\star\approx 5$ or $\lambda^\star/\langle g\rangle \approx 1$. This weak selection threshold is consistent with the cases of $c=m=10^{-2}$ in Fig.~\ref{graph r}.

For completeness, we also measured how the cooperation level changes by varying the $c$ contribution (in PGG) or $m$ requirement (in R-PGG). Our results are shown in Fig.~\ref{graph c} in Appendix~C for different topologies. The comparison supports our original observation about the equivalence of homogeneous PGG and R-PGG, no matter whether we consider spatially structured populations.

Summing up, both $r$ $\&$ $\lambda$ and $c$ $\&$ $m$ dependencies demonstrate nicely that R-PGG and PGG models are equivalent if applying the appropriate $\lambda-m$ or $r-c$ parameter pairs, no matter we have unstructured or structured populations with different interaction topologies or how intensive the selection strength is. The only crucial condition, as we will demonstrate in the next section, is that the population should be homogeneous where all players can be characterized by the same parameter value of the game.

\section{Difference between R-PGG \& PGG}
\label{difference}

Our abovementioned conclusion becomes invalid if the population is more realistic and players are not uniform anymore. Because of realistic conditions, we here focus on structured populations where interactions are limited and fixedpermanent. The introduction of heterogeneity reveals the difference between PGG and R-PGG models, as we illustrate in Sec.~\ref{hetestruc}. By means of the translation of payoff functions we will argue that R-PGG is not a simple transformation of  the original PGG, but instead a reversed form of it. 

\subsection{Evolutionary dynamics in structured populations (the heterogeneous case)}
\label{hetestruc}

To generalize our model, we here assume that players can be heterogeneous hence they may not participate in the same way in the game. This can be done by introducing player-specific parameter values. Accordingly, we assume uniform random distribution for each player's parameters $\lambda_i$ and $m_i$. For each player $i$, we randomly generate two numbers between 0 and 1: $\chi_i^{(\lambda)}\in [0,1)$, $\chi_i^{(m)}\in [0,1)$. Then, we set the parameters $\lambda_i$ and $m_i$ for player $i$ as
\begin{subequations}\label{etalambda}
	\begin{align}
	\lambda_i&=\lambda+(-2\chi_i^{(\lambda)}+1)\eta_{\lambda}\,,\\
	m_i&=m+(-2\chi_i^{(m)}+1)\eta_m\,,
	\end{align}
\end{subequations}
where $\eta_{\lambda}$ ($\eta_{\lambda}\geq 0$) measures the heterogeneity of the waste factor $\lambda_i$, and $\eta_m$ ($\eta_m\geq 0$) measures the heterogeneity of the requirement $m_i$. In this way, $\lambda_i$ and $m_i$ are selected randomly from $[\lambda-\eta_{\lambda},\lambda+\eta_{\lambda})$ and $[m-\eta_m,m+\eta_m)$ intervals where $\lambda$ and $m$ values act as the ``baselines": the expected waste factor and requirement over all players are still $\int_{0}^{1}\lambda_i~\mathrm{d}\chi_i^{(\lambda)}=\lambda$ and $\int_{0}^{1}m_i~\mathrm{d}\chi_i^{(m)}=m$. In addition, by taking $\eta_{\lambda}=0$, $\eta_m=0$, we return to the homogeneous case where $\lambda_i=\lambda$ and $m_i=m$ for all players.

For PGG, we introduce the heterogeneity in the same way. Here, the two key parameters are $\eta_r$ and $\eta_c$, which make $r$ and $c$ player-specific in the extended case. As we demonstrated in the previous section, $\lambda$ \& $r$, furthermore $m$ \& $c$ parameters are identical in the original and reversed games, therefore the $\eta_\lambda$ \& $\eta_r$ and $\eta_m$ \& $\eta_c$ amplitudes have similar roles in characterizing heterogeneity in the extended versions. 

\begin{figure}[h!]
	\centering
	\includegraphics[width=14.5cm]{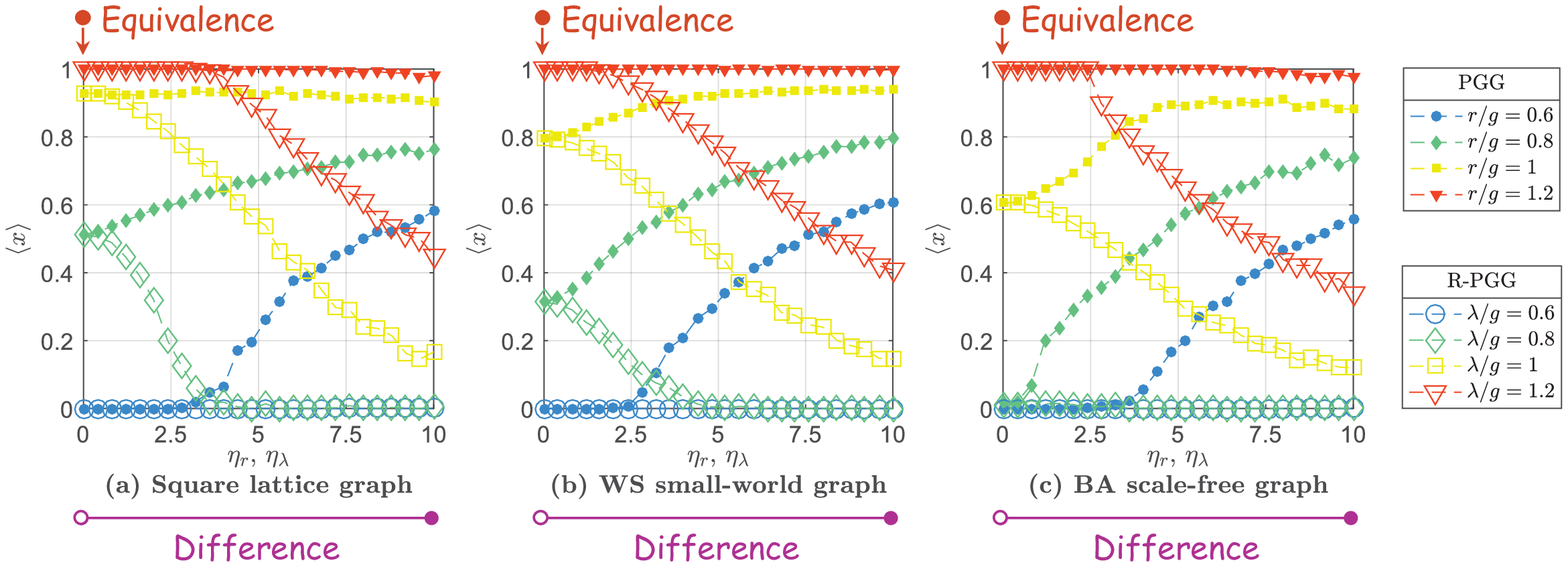}\\
	\caption{The cooperation level in dependence on the heterogeneity of the synergy factor or the waste factor in structured populations. Panels show different topologies, as indicated. When $\eta_r=0$ \& $\eta_\lambda=0$, PGG \& R-PGG are equivalent. But their difference becomes crucial as we increase $\eta_r$ and in parallel $\eta_\lambda$. We fix $c=m=1$.}\label{graph hetero r}
\end{figure}

Our key findings are summarized in Fig.~\ref{graph hetero r}, where we plot the $\langle x\rangle$ cooperation level in dependence of $\eta_r$ or $\eta_\lambda$. While they are still equivalent when $\eta_\lambda=\eta_r=0$ (in the homogeneous case), their difference becomes striking as we increase the heterogeneity of the systems. In agreement with previous observations about the principal role of heterogeneity \cite{perc_pre08,santos_n08}, the cooperation level increases as we increase $\eta_r$ in PGG for all interaction graphs. But the opposite is true for R-PGG where the extension to a heterogeneous population has the reversed consequence for the cooperation level. While increasing the heterogeneity in the synergy factor usually promotes cooperation in PGG, the opposite is observed for unequal waste factors in R-PGG. This is generally true independently of the applied topology or the dilemma strength.

Our observations can be explained as follows. In PGG, those with a higher synergy factor tend to have a higher payoff, for themselves and their co-players, which leads to an environment beneficial to cooperation. The strategy of cooperation, hence, is favored to reproduce. In R-PGG, however, if we transform the payoff function (see Eq.~(\ref{rpgg1}) for details), we can observe a redundant item $-m\lambda$ compared with the payoff function of PGG. Originally, those with a higher waste factor in R-PGG could have the chance of a higher payoff. Nevertheless, the item $-m\lambda$ tends to bring them a lower payoff. As a result, the same effect in PGG is inhibited in R-PGG, and the heterogeneity of the waste factor usually hinders cooperation in R-PGG.

Evidently, we can introduce heterogeneity in an alternative way when a cooperator's contribution (in PGG) or a defector's requirement (in R-PGG) becomes player-specific. Accordingly, we use individual $c_i$ or $m_i$ values as it is introduced in Eq.~\ref{etalambda}b. Our results are summarized in Fig.~\ref{graph hetero c} of Appendix~D. The comparison of PGG and R-PGG confirms what we found previously. Namely, these systems show similar cooperation levels only if the heterogeneity is mild. But their difference becomes striking if we increase the amplitude of $\eta_c$ and $\eta_m$.
\begin{figure}[h!]
	\centering
	\includegraphics[width=13.5cm]{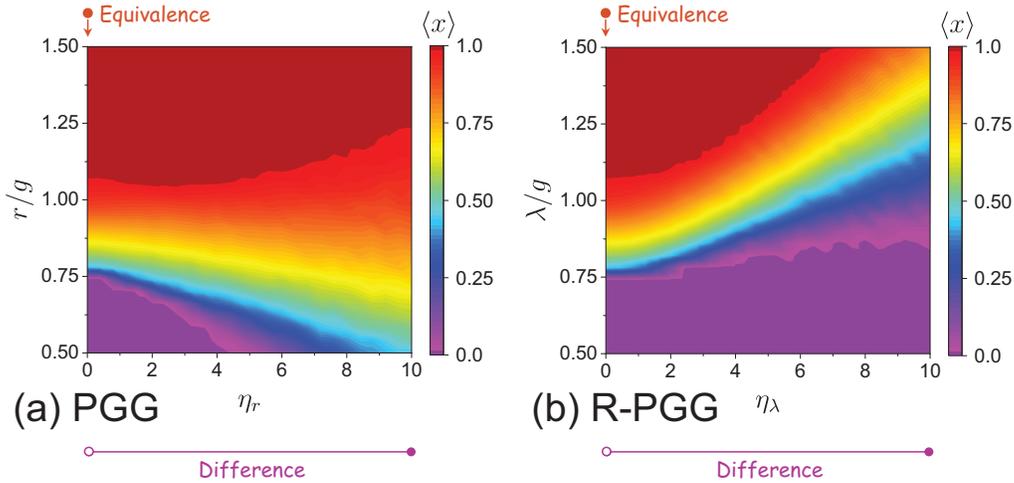}\\
	\caption{(a) The heat map of cooperation level on the parameter plane of normalized synergy factor $r/g$ and heterogeneous $r_i$. (b) The same plot in dependence of normalized waste factor $\lambda/g$ and heterogeneous $\lambda_i$. The difference between PGG and R-PGG is more striking at low $r$ and low $\lambda$ region when there is a proper social dilemma and less noticeable for high $r$ and high $\lambda$ values. But it still exists if the individual heterogeneity is large enough. The simulations are performed on square lattice topology. We fix $c=m=1$.}\label{graph r hetero r}
\end{figure}

A more comprehensive overview of the system behavior can be obtained if we plot the cooperation level not just at a specific $r$ or $\lambda$, but for arbitrarily large average values. For simplicity, we only show the results obtained on square lattice topology. The results can be seen in Fig.~\ref{graph r hetero r} where we present the cooperation level as a heat map on a two-dimensional parameter plane. Notably, Fig.~\ref{graph hetero r} can be considered a cross-section of this new diagram. Our first qualitative observation is the ``red'' area, which represents the highest cooperation level in the population, is significantly larger for PGG than for R-PGG, which highlights that heterogeneity is generally more beneficial for the PGG system. If we compare the heat maps more carefully then we can see that the difference between PGG and R-PGG is more striking in the low $r$ and low $\lambda$ region when there is a proper dilemma situation. In this parameter region, even minimal cooperation is challenging for the R-PGG model if the average $\lambda$ value remains below $\lambda/g \approx 0.75$. For large $r$ and large $\lambda$ values the difference is less visible between the models. But it still exists if the amplitude of individual heterogeneity is large enough. Hence, we can generally conclude that the heterogeneous synergy factor in PGG can promote cooperation while the heterogeneous waste factor in R-PGG cannot.

We have also compared the cooperation levels by means of heat maps when heterogeneity was introduced via individual $c_i$ and $m_i$ values. The results are shown in Fig.~\ref{graph r hetero c} in Appendix~D. The difference between PGG and R-PGG can also be revealed, showing that heterogeneity is a general condition to make a difference between these games. Summing up, we can conclude that introducing the same level of heterogeneity has significantly different consequences on PGG and R-PGG systems. This effect is robust and remains valid no matter how we introduce the distinction among players.

\subsection{Discussion on the translation of payoff function}
\label{translation}

\begin{figure}[h!]
	\centering
	\includegraphics[width=13cm]{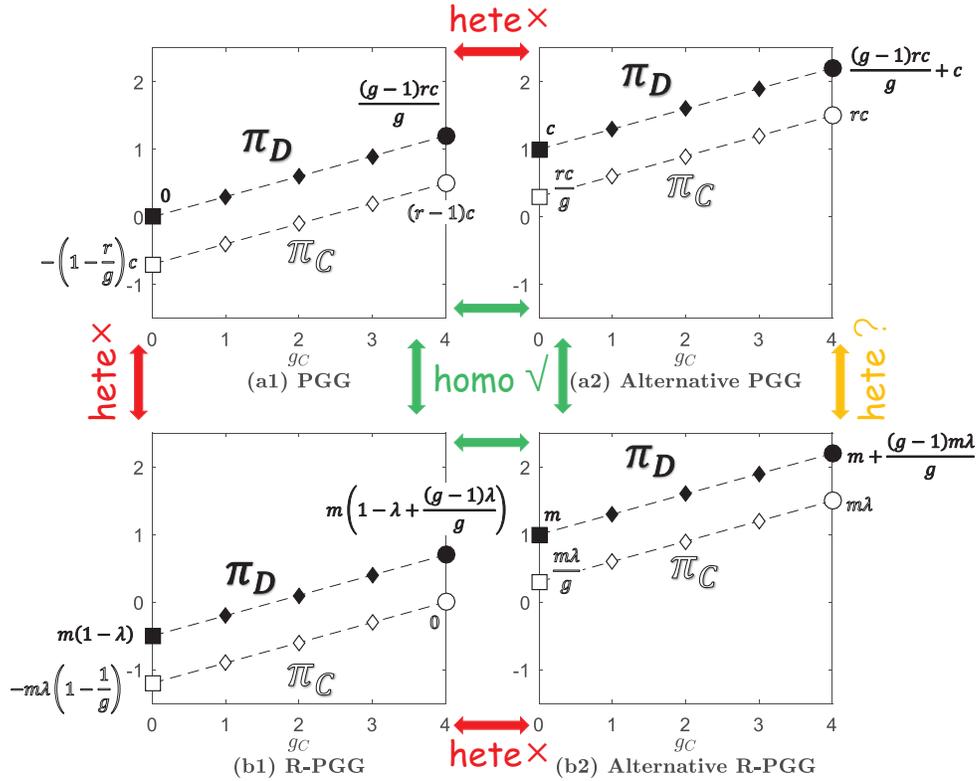}\\
	\caption{Payoff value for a focal cooperator or defector player as a function of $g_C$, denoting the number of cooperators among neighbors. The four panels show the cases as indicated in the titles. They are equivalent in the payoff gap because the translational items in $\pi_D-\pi_C$ can be subtracted. They are different in $\pi_C(g_C=0)$, $\pi_C(g_C=g-1)$, $\pi_D(g_C=0)$, and $\pi_D(g_C=g-1)$, which means that their translational items in $\pi_D-\pi_C$ cannot be subtracted when parameters become heterogeneous. The green arrows indicate that the given game models are equivalent in the homogeneous case. The red arrows mean the two models are different in the heterogeneous case. The yellow arrow means the payoff functions of the models are seemingly equivalent, but later we will show they are actually different. $g=5$, $r=\lambda=1.5$, and $c=m=1$ parameters were used.}\label{figureasymmetric}
\end{figure}

To explain why R-PGG and PGG become different when heterogeneous players are present, we study the translation of payoff functions. 
As Equation~(\ref{payoffCD}) describes, a focal player's income from a group venture can always be expressed as a function of $g_C$ which is the number of cooperators among group neighbors. In the following, we consider four different translations of PGG. The details of these games are as follows.
\begin{itemize}
\item (i) {\bf PGG}
In the most well-known and widely accepted version of PGG a cooperator contributes a $c$ amount to the common pool, while a defector player does not. The sum of contributions is enlarged and distributed among all group members. The corresponding payoff values of the focal player playing either $C$ or $D$ in the group are
\begin{align}\label{pgg1}
	\begin{cases}
		\displaystyle{
		\pi_C=\frac{(g_C+1) rc}{g}-c}\\
		\displaystyle{
		\pi_D=\frac{g_C rc}{g} } \,.
	\end{cases}
\end{align}

\item (ii) {\bf Alternative PGG}
In an alternative form, all players have an initial endowment $c$ \cite{hauser2019social}. A cooperator player invests this whole amount to the common pool, while a defector player keeps it. As previously, the contributions are summed, enhanced and distributed among all competitors. The resulting payoff values are 
\begin{align}\label{pgg2}
	\begin{cases}
		\displaystyle{
			\pi_C=\frac{(g_C+1) rc}{g}}\\
		\displaystyle{
			\pi_D=\frac{g_C rc}{g}+c } \,.
	\end{cases}
\end{align}

\item (iii) {\bf R-PGG}
According to our proposal, in an R-PGG a defector requires $m$ from a common resource, while a cooperator player does not. The sum of required goods is wasted at a certain rate, which means an enlarged, and equally shared cost to everyone in the group, yielding the payoff values 
\begin{align}\label{rpgg1}
	\begin{cases}
		\displaystyle{
			\pi_C=-\frac{g_D m\lambda}{g}}\\
		\displaystyle{
			\pi_D=m-\frac{(g_D+1) m\lambda}{g} }
	\end{cases}
		\stackrel{g_C+g_D=g-1}{\Longleftrightarrow}
	\begin{cases}
		\displaystyle{
			\pi_C=-m\lambda +\frac{(g_C+1) m\lambda}{g}}\\
		\displaystyle{
			\pi_D=-m\lambda +m+\frac{g_C m\lambda}{g} } \,.
	\end{cases}
\end{align}

\item (iv) {\bf Alternative R-PGG}
In the alternative version, each player has an initial $m\lambda$ deposit. Additionally, a defector requires $m$ from a common resource, and related (enlarged) cost is shared among everyone in the group. Therefore, the payoff values are 
\begin{align}\label{rpgg2}
	\begin{cases}
		\displaystyle{
			\pi_C=m\lambda-\frac{g_D m\lambda}{g}}\\
		\displaystyle{
			\pi_D=m\lambda+m-\frac{(g_D+1) m\lambda}{g} }
	\end{cases}
	\stackrel{g_C+g_D=g-1}{\Longleftrightarrow}
	\begin{cases}
		\displaystyle{
			\pi_C=\frac{(g_C+1) m\lambda}{g}}\\
		\displaystyle{
			\pi_D=m+\frac{g_C m\lambda}{g} }\,.
	\end{cases}
\end{align}
\end{itemize}
Note that in Eq.~(\ref{rpgg1}) and in Eq.~(\ref{rpgg2}) we have used $g_D=g-g_C-1$ notation to obtain a consistent $g_C$-dependent form of payoff values for all cases. The connections between the above-specified cases are summarized in Fig.~\ref{figureasymmetric}, where the curves represent the payoff of a focal player employing either cooperation or defection.

The key assumption, which explains the equivalence, is the linear dependence on the $g_C$ value. We can see that with the same parameters, the slope of payoff functions is constant everywhere. Also, the difference between $\pi_D$ and $\pi_C$ (i.e., $\pi_D-\pi_C$) is constant at any $g_C$ value. Therefore, despite different interpretations, we can say that all four translations expressed in panels~(a1),~(a2),~(b1), and (b2) represent conceptually similar public goods games. 

The equivalence in the homogeneous case is clear because the results depend only on $\pi_{i'}-\pi_i$. Here the translational items in $\pi_{i'}-\pi_i$ can be subtracted, which leads to the equivalence among the four interpretations. 
However, the absolute values of payoff functions vary from case to case, as indicated by the value of $\pi_C(g_C=0)$, $\pi_C(g_C=g-1)$, $\pi_D(g_C=0)$, and $\pi_D(g_C=g-1)$ in Fig.~\ref{figureasymmetric}. This leads to a significant difference when introducing heterogeneity. As an example, take alternative PGG and original R-PGG versions. We can see that R-PGG has a redundant item $-m\lambda$ compared with alternative PGG. As mentioned earlier, in a homogeneous case, such an item can be subtracted when calculating $\pi_{i'}-\pi_i$ for strategy update. However, in a heterogeneous case, such an item varies from player to player, $-m_i\lambda_i\neq -m_{i'}\lambda_{i'}$, which cannot be subtracted anymore. 
\begin{figure}[h!]
	\centering
	\includegraphics[width=13cm]{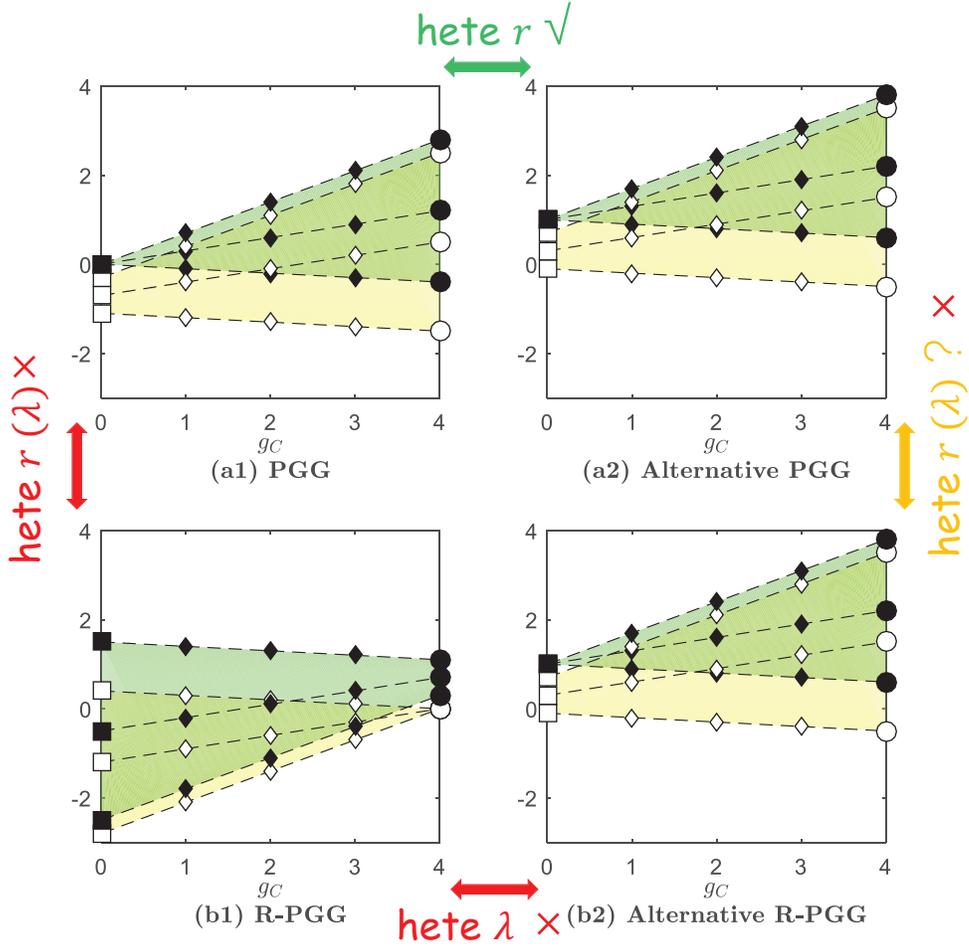}\\
	\caption{Payoff values for a cooperator (open symbols) and for a defector (solid symbols) player in dependence on the number of other cooperative players in the group. Panels show different models, as indicated. While $g=5$, $c=m=1$ values are fixed, the three curves for each $\pi_C$ or $\pi_D$ are plotted by taking $r_i=\lambda_i=1.5-2$, $1.5$, $1.5+2$ ($r=\lambda=1.5$, $\eta_r=\eta_{\lambda}=2$). The shaded areas mark the traces of curves when $r_i$ or $\lambda_i$ changes in the mentioned interval bordered by $r\pm\eta_r$ ($\lambda\pm\eta_{\lambda}$). These areas are marked by different colors for competing strategies (green for $D$ and yellow for $C$). The meaning of color arrows are similar to those we used in Fig.~\ref{figureasymmetric}.}\label{figureasymmetric heter}
\end{figure}

To visualize this effect in a more intuitive way, we present Fig.~\ref{figureasymmetric heter}  which shows $\pi_C$ and $\pi_D$ as a function of $g_C$ in the mentioned cases when heterogeneous $r$ or $\lambda$ are introduced. The curves represent three specific average values of key parameters, but we also mark the location of payoff values for intermediate values by shaded areas. In this way the conceptual difference between different cases becomes clear. Staying at Fig.~\ref{figureasymmetric heter}, the shaded areas in panel~(a1) and in panel~(b1) are completely different because the difference in the payoff values leads to unequal outcomes in the presence of heterogeneity. 

Notably, the shape of the shaded area remains intact if we change PGG, shown in panel Fig.~\ref{figureasymmetric heter}(a1), to the alternative PGG, presented in panel Fig.~\ref{figureasymmetric heter}(a2). This is because their coefficients in front of $r$ are equal: $(g_C+1)c/g$ in $\pi_C$, and $g_C c/g$ in $\pi_D$. Therefore, the same change in $r$ leads to the same change in the payoff function. Although they are still different in their vertical positions, the identical shape confirms their equivalence in the stationary cooperation level (the reader may compare Fig.~\ref{figurealter r}(a) in Appendix~D and Fig.~\ref{graph r hetero r}(a)). Furthermore, if we compare the case of alternative PGG (Fig.~\ref{figureasymmetric heter}(a2)) and the case of alternative R-PGG (Fig.~\ref{figureasymmetric heter}(b2)) then we can see that they show not only the same shape of shaded areas, but also the same absolute payoff values. Their cooperation levels, however, as we  show in Fig.~\ref{figurealter r} in Appendix~D, are different. This fact is a direct consequence of the conceptual difference between PGG $\&$ R-PGG, as we will discuss later.

For completeness, we also studied the consequence of heterogeneity in $c$ or $m$ parameter on the payoff functions. The results are presented in Fig.~\ref{figureasymmetric hetec} in Appendix~D. It shows that the same level of change in individual $c$ leads to a varying change in their payoff functions, resulting in different shapes of shaded areas. We, however, stress that alone the payoff functions are not able to reveal all differences among the cases we discussed above. For example, similar shapes of shaded areas can produce unequal heat maps of cooperation levels on the parameter plane, as shown in Appendix~D.

To give a deeper insight into the origin of varying behavior in heterogeneous games, let us turn back to the payoff functions. When we used the term $g_C=g-g_D-1$ to replace $g_D$ with $g_C$ in R-PGG (as well as in alternative R-PGG), we implicitly assumed homogeneous payoff functions. Therefore, the individual parameter values have no importance. However, if we assume distinct players and consider player-specific payoff values via Eq.~(\ref{unequalpayoff}), then we can rewrite the payoff function for the heterogeneous cases. For example, the payoff for the original PGG can be written as
\begin{equation}
	\pi_i=\frac{1}{g}\sum_{j=1}^g x_j c_j r_j-x_i c_i\,,
\end{equation}
and for the original R-PGG it is
\begin{align} \label{transRPGG}
	\pi_i&=y_i m_i-\frac{1}{g}\sum_{j=1}^g y_j m_j\lambda_j\nonumber\\
	&=(1-x_i) m_i-\frac{1}{g}\sum_{j=1}^g (1-x_j) m_j\lambda_j\nonumber\\
	&=-\frac{\sum_{j=1}^g m_j\lambda_j}{g}+\frac{1}{g}\sum_{j=1}^g x_j m_j\lambda_j+(1-x_i) m_i\,.
\end{align}
Now we can see that the term $-\sum_{j=1}^g m_j\lambda_j/g$ is related to the waste factor of all players in the group. Importantly, we can write the payoff for alternative R-PGG as
\begin{align} \label{transalterRPGG}
	\pi_i&=m_i\lambda_i+y_i m_i-\frac{1}{g}\sum_{j=1}^g y_j m_j\lambda_j\nonumber\\
	&=m_i\lambda_i+(1-x_i) m_i-\frac{1}{g}\sum_{j=1}^g (1-x_j) m_j\lambda_j\nonumber\\
	&=m_i\lambda_i-\frac{\sum_{j=1}^g m_j\lambda_j}{g}+\frac{1}{g}\sum_{j=1}^g x_j m_j\lambda_j+(1-x_i) m_i\,,
\end{align}
while for alternative PGG it is
\begin{equation} \label{alterPGG}
	\pi_i=\frac{1}{g}\sum_{j=1}^g x_j c_j r_j+(1-x_i) c_i\,.
\end{equation}

By comparing Eq.~(\ref{transalterRPGG}) and Eq.~(\ref{alterPGG}), we can identify a non-zero redundant item $m_i\lambda_i-(\sum_{j=1}^g m_j\lambda_j)/g$ in R-PGG. This is the reason behind the observed difference in cooperation levels when we change the independent variable from $g_D$ to $g_C$. In particular, if we apply identical extension to PGG and alternative PGG, then we have no such difficulties because cooperators take the key role hence crucial game parameters belonging to this strategy in both model versions. A similar argument can be raised when we compare R-PGG and alternative R-PGG, because the decisive game parameters determine the payoff values that belong to defectors in both models. This explains why we cannot transform the family of PGGs into the family of R-PGGs by simply switching cooperators \& defectors. 

The conceptual difference between the models can be termed a ``reversed effect'' and its essence is explained in the following way. The game parameters working in PGG \& R-PGG belong to different strategies. In PGG, the personal features of cooperator players count: if a player $i$ cooperates, she contributes $c_i$ and enlarges the contribution by $r_i$; meanwhile, a defective player $j$ has no chance to ``activate'' her specific $c_j$ and $r_j$ values. But she only receives the public goods of $r_i c_i$ produced by a cooperator. On the contrary, in R-PGG, the diversity of defectors players becomes essential: if a player $j$ defects, she receives $m_j$ which is multiplied by $\lambda_j$ factor as the public cost. Meanwhile, a cooperative player $i$ never has a chance to validate her $m_i$ and $\lambda_i$ parameter values. Her role is limited to sharing the public cost of $m_j \lambda_j$ of a defector player.

In this way, the role of defectors is essential in R-PGG which is irreplaceable in a more complex environment. Hence, we can conclude that R-PGG is not a simple translational transformation of the original PGG, but rather a reversed form of it.
 
\section{Conclusion}
\label{conclusion}

Generally, people make logical, and frequently economical decisions when their consumption and its cost directly relate. However, if this connection is less clear, because the emerging cost is shared among a group of people, then we face a dilemma situation: we may require more goods from a common source than it is absolutely necessary because the related extra expense will be covered collectively. The reverse is also true: our economic behavior by consuming fewer resources will not necessarily be awarded by a proportionally smaller bill in the mentioned situation because all the others will also benefit from it. One might think this social dilemma is just a simple transformation of the original PGG where players contribute freely to a common pool and receive goods equally. In the reversed case described above, players receive goods from a common resource freely and should contribute equally. But the relation between the original PGG and our present R-PGG is more subtle.

In the case of a homogeneous population, where potential contributions and demands are constant, R-PGG is equivalent to PGG. This fact can be supported by means of a static calculation, deterministic and stochastic evolutionary analysis in a well-mixed population, and Monte Carlo simulations in structured populations. In evolutionary game dynamics, the requirement $m$ parameter in R-PGG plays the same role as the cooperator contribution $c$ does in PGG. Furthermore, the waste factor $\lambda$ in R-PGG has the same duty as the synergy factor $r$ has in PGG. By transforming these parameters between R-PGG and PGG, we can obtain exactly the same system behaviors.

The network reciprocity reveals the following conclusions in R-PGG. If requiring goods leads to more cost to the group, then players tend to require fewer goods. Intriguingly, if players are allowed to require more goods, then fewer players may do it. The opposite behavior, which is consistent with our intuition, can only be detected clearly on small-world and scale-free graphs under the weak dilemma limit.

Although the underlying R-PGG model seems equivalent to PGG, the agreement diminishes when we introduce more realistic conditions, like supposing heterogeneous players with diverse contribution capacities or unequal requirement levels for common resources. In the mentioned cases, R-PGG and PGG show major differences and produce highly different cooperation levels even if we apply the same average values of key parameters with equally strong amplitudes of diversity. The key observation is that while heterogeneity can generally support cooperation in PGG, the mentioned condition impedes the evolution of cooperation in R-PGG.

To understand the deeper origin of these diverse behaviors, we studied how payoff functions vary in response to the heterogeneity condition for different cases. In the case of equivalent models, the mentioned function should vary similarly if we apply varying average or growing diversity of key parameters. This requirement is justified between alternative forms of the original PGG where cooperators play the decisive role in collective income. This is also true between the alternative versions of R-PGG where the individual defector's profile becomes crucial. In this way, the equivalence cannot be held between PGG and R-PGG where the heterogeneity reveals the conceptual difference between the model families. Hence, we can say that R-PGG is not a trivial transformation of the original PGG, but can be considered a reversed version of it.

Our present work focused on the difference in structured populations, but of course, similar study can also be done in a well-mixed system. Hopefully, our observations will stimulate forthcoming research. When the applied microscopic rule leads to the heterogeneity of game parameters by any means, one can study the consequence of extended parameters on PGGs and R-PGGs separately. The different results under the same topic (e.g., environmental protection) can reveal how the introduced rule works in PGGs (e.g., players donate to environmental protection), and how it works in corresponding R-PGGs (e.g., players litter for convenience). Aside from heterogeneity, the potential difference between R-PGG and PGG can be further investigated.

\section*{Acknowledgements}
A.S. was supported by the National Research, Development and Innovation Office (NKFIH) under Grant No. K142948.

\appendix
\section{Equivalence in well-mixed populations}\label{A}
\subsubsection*{A1. Deterministic dynamics}

According to the replicator equation approach \cite{schuster1983replicator}, we denote the frequency of cooperation in the population by $f_C$ and the frequency of defection by $f_D$, yielding $f_C+f_D=1$. In an R-PGG, a player is selected to form a group with $g-1$ other randomly chosen players. The probability that there are $g_C$ cooperators and $g_D$ defectors among the neighbors can be given as $\sum_{g_D=0}^{g-1}\dbinom{g-1}{g_D}{f_D}^{g_D}{f_C}^{g_C}$.

Therefore, based on the payoff definition given by Eq.~(\ref{payoffC}) and Eq.~(\ref{payoffD}), the expected payoff $\langle \pi_C\rangle$ and $\langle \pi_D\rangle$ for the selected cooperator or defector player is
\begin{subequations}
	\begin{align}
		\langle \pi_C\rangle&=\sum_{g_D=0}^{g-1}\dbinom{g-1}{g_D}{f_D}^{g_D}(1-f_D)^{g-g_D-1}\pi_C=-\frac{m\lambda(g-1)}{g}f_D \label{deterexpayc}\\
		\langle \pi_D\rangle&=\sum_{g_D=0}^{g-1}\dbinom{g-1}{g_D}{f_D}^{g_D}(1-f_D)^{g-g_D-1}\pi_D=m-\frac{m\lambda(g-1)}{g}f_D-\frac{m\lambda}{g} \label{deterexpayd}\,.
	\end{align}
\end{subequations}

By using these values, the time-derivative of the frequencies of competing strategies are
\begin{subequations}\label{deterfcfd}
	\begin{align}
		\dot f_C &= f_Cf_D Q(\langle \pi_D\rangle \gets \langle \pi_C\rangle)-f_Df_C \,Q(\langle \pi_C\rangle \gets \langle \pi_D\rangle)\nonumber \\
		&=f_C(1-f_C)\left( \frac{1}{1+\e^{-\omega (\langle \pi_C\rangle-\langle \pi_D\rangle)}}-\frac{1}{1+\e^{-\omega (\langle \pi_D\rangle-\langle \pi_C\rangle)}}\right) \label{deterfc}\\
		\dot f_D &=-\dot f_C \,. \label{deterfd}
	\end{align}
\end{subequations}

The stationary solutions are ${f_C}^*=0$ and ${f_C}^*=1$ whose stability depends only on the sign of $\langle \pi_D\rangle-\langle \pi_C\rangle$:
\begin{equation}\label{deterpaygap}
	\langle \pi_D\rangle-\langle \pi_C\rangle=m(1-\frac{\lambda}{g})\,.
\end{equation}
Accordingly, if $\lambda<g$, then $\langle \pi_D\rangle>\langle \pi_C\rangle$ and ${f_C}^*=0$ is stable, while for $\lambda>g$, ${f_C}^*=1$ is stable. Intuitively, if requiring goods brings too much waste and loss for the group, then players prefer not requiring and being cooperative.

For the related PGG (see Eq.~(\ref{pgg1})), we know $\langle \pi_D\rangle-\langle \pi_C\rangle = (1-r/g)c$. In this way, $\lambda$ \& $m$ parameters play exactly the same role in R-PGG as $r$ \& $c$ parameters in the original PGG. Hence, we can conclude that the properties of homogeneous R-PGG are equivalent to the behaviors of homogeneous PGG in the framework of replicator dynamics in a well-mixed infinite population.

\subsubsection*{A.2. Stochastic dynamics}

Next, we consider a finite population of size $N$, in which $N_C$ cooperators are present. For a cooperative focal player, the expected payoff $\langle \pi_C\rangle$ is
\begin{subequations}\label{stoexpa}
	\begin{align}
		\langle \pi_C\rangle=\sum_{g_D=0}^{g-1}\dfrac{\dbinom{N-N_C}{g_D}\dbinom{N_C-1}{g-g_D-1}}{\dbinom{N-1}{g-1}}\pi_C=-\frac{m\lambda (g-1)(N-N_C)}{g(N-1)}
		\tag{\ref{stoexpa}{a}} \label{stoexpac}\,.
	\end{align}
\end{subequations}
Similarly, the expected payoff $\langle \pi_D\rangle$ for a defective focal player is
\begin{align}
	\langle \pi_D\rangle=\sum_{g_D=0}^{g-1}\dfrac{\dbinom{N-N_C-1}{g_D}\dbinom{N_C}{g-g_D-1}}{\dbinom{N-1}{g-1}}\pi_D=m-\frac{m\lambda (g-1)(N-N_C)}{g(N-1)}-\frac{m\lambda (N-g)}{g(N-1)}
	\tag{\ref{stoexpa}{b}}. \label{stoexpad}
\end{align}
Therefore, we have
\begin{equation}\label{stopaygap}
	\langle \pi_D\rangle-\langle \pi_C\rangle=m\left(1-\frac{\lambda}{g}\frac{N-g}{N-1}\right)\,.
\end{equation}

If we denote by $P_{N_C}^+$ ($P_{N_C}^-$) the transition probability that $N_C$ is increased (decreased) by 1 then 
\begin{subequations}\label{transi}
	\begin{align}
		P_{N_C}^+=\frac{N-N_C}{N}\frac{N_C}{N}\,Q(\langle \pi_D\rangle \gets \langle \pi_C\rangle)=\frac{N-N_C}{N}\frac{N_C}{N}\frac{1}{1+\e^{-\omega (\langle \pi_C\rangle-\langle \pi_D\rangle)}}
		\tag{\ref{transi}{a}} \label{transi+}
	\end{align}
\end{subequations}
and
\begin{align}
	P_{N_C}^-=\frac{N_C}{N}\frac{N-N_C}{N}\,Q(\langle \pi_C\rangle \gets \langle \pi_D\rangle)=\frac{N_C}{N}\frac{N-N_C}{N}\frac{1}{1+\e^{-\omega (\langle \pi_D\rangle-\langle \pi_C\rangle)}}
	\tag{\ref{transi}{b}} \label{transi-}\,.
\end{align}
Consequently, the $P_{N_C}^0$ probability that $N_C$ keeps unchanged is
\begin{align}
	P_{N_C}^0=1-P_{N_C}^+-P_{N_C}^-
	\tag{\ref{transi}{c}} \label{transi0}\,.
\end{align}

At time step $t$, the probability that the system is found in state $N_C$ is described by $\varphi_{N_C}(t)$. The master equation of $\varphi_{N_C}(t)$ is
\begin{equation}\label{mastervarphi}
	\varphi_{N_C}(t+1)=P_{N_C}^0\varphi_{N_C}(t)+P_{N_C}^-\varphi_{N_C+1}(t)+P_{N_C}^+\varphi_{N_C-1}(t)\,.
\end{equation}

The system has two absorbing states: $N_C=0$ and $N_C=N$, in which $P_{N_C}^-=P_{N_C}^+=0$, $P_{N_C}^0=1$, and $\varphi_{N_C}(t+1)=\varphi_{N_C}(t)$. 

The fixation probability that initial $N_C$ cooperative players finally take over the whole population is denoted by $\rho_{N_C}$. In previous works \cite{nowak2004emergence,karlin2014first}, the master equation of $\rho_{N_C}$ has been deduced:
\begin{equation}\label{fixp}
	\rho_{N_C}=\frac{\sum_{i=0}^{N_C-1}\prod_{j=1}^{i}\dfrac{P_j^-}{P_j^+}}{\sum_{i=0}^{N-1}\prod_{j=1}^{i}\dfrac{P_j^-}{P_j^+}}\,.
\end{equation}

By taking $N_C=1$, we have the probability $\rho_1$ that a single cooperative player can invade and take over the remaining $N-1$ defective players,
\begin{equation}\label{fixp1}
	\rho_1=\frac{1}{1+\sum_{i=1}^{N-1}\prod_{j=1}^{i}\dfrac{P_j^-}{P_j^+}}\,.
\end{equation}

If we suppose neutral selection (i.e., $\omega=0$), we have $P_{N_C}^-/P_{N_C}^+=1$, such that $\rho_{N_C}=N_C/N$ and $\rho_{1}=1/N$. Furthermore, if we consider weak selection (i.e., $0<\omega \ll 1$), we can use Taylor expansion to approximate the expression,
\begin{equation}\label{weakp-p+}
	\frac{P_j^-}{P_j^+}=\mathrm{e}^{-\omega (\langle \pi_C\rangle-\langle \pi_D\rangle)}\approx 1-\omega (\langle \pi_C\rangle-\langle \pi_D\rangle) = 1+\omega m\left(1-\frac{\lambda}{g}\frac{N-g}{N-1}\right)\,.
\end{equation}

Substituting Eq.~(\ref{weakp-p+}) into Eq.~(\ref{fixp1}) leads to the fixation probability
\begin{equation}\label{fixp1value}
	\rho_1=-\frac{\omega m\left(1-\dfrac{\lambda}{g}\dfrac{N-g}{N-1}\right)}{1-\left(1+\omega m\left(1-\dfrac{\lambda}{g}\dfrac{N-g}{N-1}\right)\right)^N}\,.
\end{equation}
Note that $\lim_{\omega \to 0}\rho_1=1/N$ in Eq.~(\ref{fixp1value}), which is consistent with the result of neutral selection. Accordingly, if $\lambda<g(N-1)/(N-g)$, then $\langle \pi_D\rangle>\langle \pi_C\rangle$ and $\rho_1<1/N$ (evolution disfavors cooperation), while for $\lambda>g(N-1)/(N-g)$, $\rho_1>1/N$ (evolution favors cooperation).

Comparing the fixation probability $\rho_1$ in R-PGG and the original PGG, we can see the parallel between parameters $\lambda$ \& $r$ and parameters $m$ \& $c$. 

\section{Threshold for cooperator success under weak selection}\label{B}

By means of the identity-by-descent method, Su {\it et al.} \cite{su2018understanding,su2019spatial} identified the threshold value of the synergy factor when natural selection favors cooperation over defection under a weak selection limit in the original PGG. Their theory was built on three assumptions: (1) transitive graph; (2) weak selection; and (3) low mutation. At a $0\leq \mu\leq 1$ mutation rate, a player randomly updates its strategy; otherwise, the player updates its strategy via a Fermi process. In this work we do not consider mutation, but only use the Fermi process to update strategies, which leads to the approximation of a low mutation $\mu\to0$ limit. In the homogeneous case, we here show that R-PGG has exactly the same $\lambda^\star$ threshold value for cooperation, as it has for the original PGG.

Based on the above assumptions and according to \cite{allen2014games}, the condition that favors cooperation is
\begin{equation}\label{bd}
	\left\langle\frac{\partial}{\partial\omega}(b_i-d_i)\right\rangle_{\begin{smallmatrix}\omega=0\\y_i=0\end{smallmatrix}}>0\,,
\end{equation}
where $b_i$ is the probability that player $i$ reproduces its strategy to a neighbor, $d_i$ is the probability that player $i$'s strategy is replaced by the strategy of a neighbor, and bracket $\langle \cdot \rangle_{\begin{smallmatrix}\omega=0\\y_i=0\end{smallmatrix}}$ denotes the average over stationary distribution given neural selection and a cooperator player $i$.

To describe the stochastic process, we define an $(n,s)$-random walk, which means that the player walks with $n$ steps on the interaction graph (where players play games) and $s$ steps on the dispersal graph (where players update strategies) \cite{allen2014games}. In this work, the interaction graph and dispersal graph overlap. We denote by $y_i^{(n,s)}$ the probability that the player at the end of player $i$'s $(n,s)$-random walk is a defector. Naturally, the probability that a neighbor of player $i$ on the dispersal graph defects is $y_i^{(0,1)}$. 

Then, for the Fermi process, $b_i$ and $d_i$ respectively are
\begin{subequations}\label{bidi}
	\begin{align}
		b_i&=\frac{1}{N}\sum_{l_{ii'}\in\mathcal{L}}\frac{Q(\pi_{i'}\gets\pi_i)}{k}
		\tag{\ref{bidi}{a}} \label{bi}\\
		d_i&=\frac{1}{N}\sum_{l_{ii'}\in\mathcal{L}}\frac{Q(\pi_i\gets\pi_{i'})}{k}
		\tag{\ref{bidi}{b}}\,. \label{di}
	\end{align}
\end{subequations}
By substituting Eqs.~(\ref{bi}) and (\ref{di}) into Eq.~(\ref{bd}), we have
\begin{align}\label{fermibd}
	&\left\langle\frac{\partial}{\partial\omega}(b_i-d_i)\right\rangle_{\begin{smallmatrix}\omega=0\\y_i=0\end{smallmatrix}}>0\nonumber\\
	\Leftrightarrow & \left\langle\frac{\partial}{\partial\omega}\left(
	\frac{1}{N}\sum_{l_{ii'}\in\mathcal{L}}\frac{1}{k}\dfrac{1}{1+\mathrm{e}^{-\omega(\pi_i-\pi_{i'})}}
	-
	\frac{1}{N}\sum_{l_{ii'}\in\mathcal{L}}\frac{1}{k}\dfrac{1}{1+\mathrm{e}^{-\omega(\pi_{i'}-\pi_i)}}
	\right)\right\rangle_{\begin{smallmatrix}\omega=0\\y_i=0\end{smallmatrix}}>0\nonumber\\
	\Leftrightarrow &
	\left\langle\pi_i\right\rangle_{\begin{smallmatrix}\omega=0\\y_i=0\end{smallmatrix}}
	-
	\frac{1}{k}\left\langle
	\sum_{l_{ii'}\in\mathcal{L}}\pi_{i'}
	\right\rangle_{\begin{smallmatrix}\omega=0\\y_i=0\end{smallmatrix}}>0\nonumber\\
	\Leftrightarrow &~
	\pi_i^{(0,0)}>\pi_i^{(0,1)}\,,
\end{align}
where the average payoff of the players at the end of $(n,s)$-random walk from player $i$ is denoted by $\pi_i^{(n,s)}$. The condition $\pi_i^{(0,0)}>\pi_i^{(0,1)}$ means that the payoff of player $i$ exceeds the average payoff of its neighbors.

According to \cite{allen2014games}, we have the following equation in the low mutation limit:
\begin{equation}\label{weakmutation}
	\lim_{\mu\to 0}(y_i^{(n,s)}-y_i^{(n,s+1)})=-\frac{\mu}{2}(Np_i^{(n,s)}-1)\,,
\end{equation}
where $p_i^{(n,s)}$ is the probability of arriving at the starting node $i$ after randomly walking $n$ steps on the interaction graph and $s$ steps on the dispersal graph from $i$.

By still using the homogeneous $m$ and $\lambda$ assumption, the expected payoff of player $i$ at the end of $(n,s)$-random walk can be calculated as
\begin{align}\label{randomwalkpayoff}
	\pi_i^{(n,s)}=&~\frac{1}{k+1}\left[ y_i^{(n,s)}m-\frac{1}{k+1}\left(y_i^{(n,s)}m\lambda+ky_i^{(n+1,s)}m\lambda\right)\right.\nonumber
	\\
	&\left.+k\left(y_i^{(n,s)}m-\frac{1}{k+1}\left(y_i^{(n+1,s)}m\lambda+ky_i^{(n+2,s)}m\lambda\right)\right)\right]\nonumber\\
	=&-\frac{k^2}{(k+1)^2}m\lambda y_i^{(n+2,s)}-\frac{2k}{(k+1)^2}m\lambda y_i^{(n+1,s)}-\frac{1}{(k+1)^2}m\lambda y_i^{(n,s)}+my_i^{(n,s)}\,.
\end{align}

If we substitute Eqs.~(\ref{randomwalkpayoff}) and (\ref{weakmutation}) into Eq.~(\ref{fermibd}), we have
\begin{align}\label{emergelimit}
	&~\pi_i^{(0,0)}>\pi_i^{(0,1)}\nonumber\\
	\Leftrightarrow &-\frac{k^2}{(k+1)^2}m\lambda (y_i^{(2,0)}-y_i^{(2,1)})-\frac{2k}{(k+1)^2}m\lambda (y_i^{(1,0)}-y_i^{(1,1)})\nonumber\\
	&-\frac{1}{(k+1)^2}m\lambda (y_i^{(0,0)}-y_i^{(0,1)})+m(y_i^{(0,0)}-y_i^{(0,1)})>0\nonumber\\
	\Leftrightarrow &-(Np_i^{(0,0)}-1)+\lambda\left[\frac{N}{(k+1)^2}\left(k^2p_i^{(2,0)}+2kp_i^{(1,0)}+p_i^{(0,0)}\right)-1\right]>0\nonumber\\
	\Leftrightarrow &~\lambda>\frac{(N-1)(k+1)}{N-(k+1)}\equiv \lambda^\star\,,
\end{align}
which gives the explicit threshold $\lambda^\star$ over which cooperation emerges under weak selection. In Eq.~(\ref{emergelimit}), we use $p_i^{(0,0)}=1$, one stays at the starting node if not walking; $p_i^{(1,0)}=0$, one cannot leave and go back to the starting node within a single step; $p_i^{(2,0)}=1/k$, the result produced by similar logic as above. Note that transitiveness unifies $p_i^{(n,s)}$ among different node $i$.
Since $g=k+1$, we have $\lambda^{\star}=g(N-1)/(N-g)$, the same value by which we judge whether evolution favors cooperation in a finite unstructured population.
We can see that the resulting $\lambda^\star$ is exactly the same as $r^\star$ obtained for the original PGG (Refs.~\cite{su2018understanding,su2019spatial}), signaling the equivalence of homogeneous R-PGG \& PGG in the weak selection limit.

\section{Monte Carlo results for homogeneous spatial populations}\label{C}

\begin{figure}[h!]
	\centering
	\includegraphics[width=14.5cm]{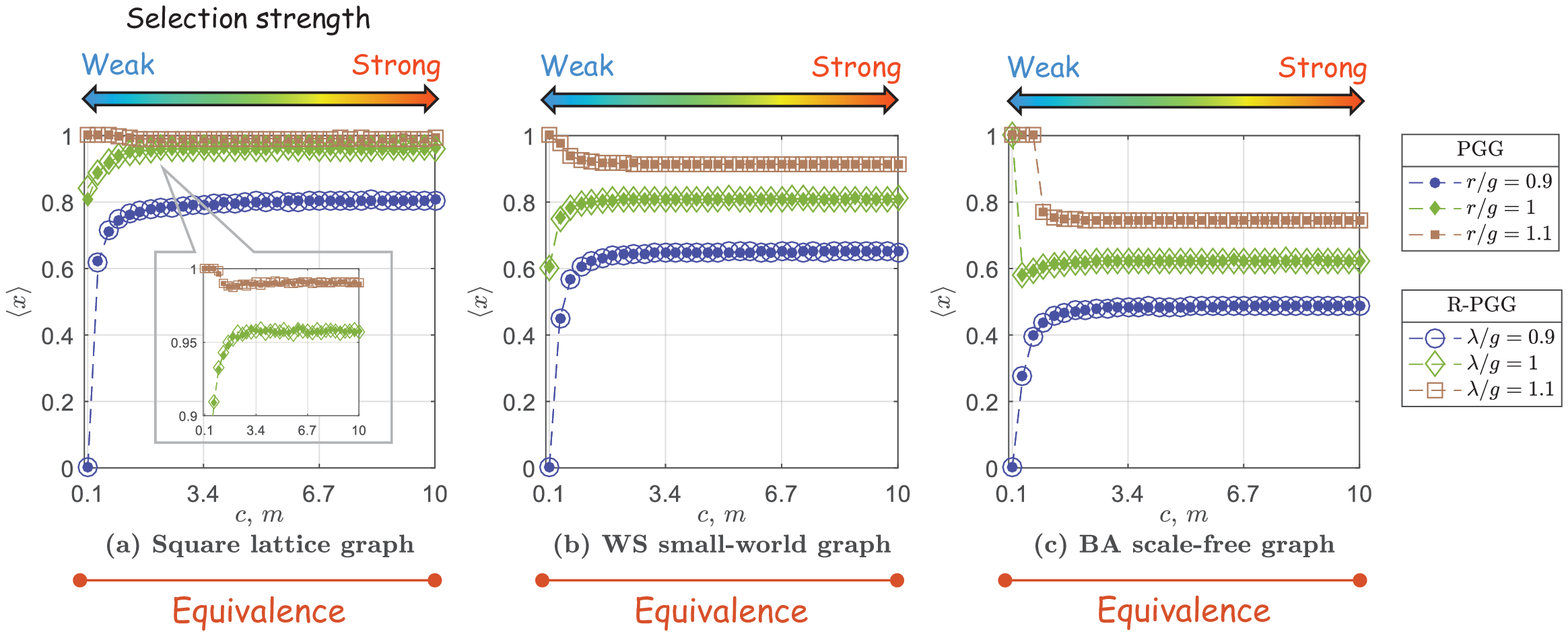}\\
	\caption{Cooperation level as a function of contribution $c$ in PGG and requirement $m$ in R-PGG at a fixed value of social dilemma strength. The normalized values of $r$ and $\lambda$ are marked by the legends. Panels show the results for different topologies. The inset in panel~(a) is the enlarged part of the upper region. On the top, we also mark how selection strength changes by $m$.}\label{graph c}
\end{figure}

Figure~\ref{graph c} presents MC results for average $\langle x\rangle$ cooperation level as a function of $c$ (in PGG) or $m$ (in R-PGG) for different interaction topologies. The details of the simulations agree with those specified in the main text. It may seem puzzling that in R-PGG the increase in consumption range may result in a higher cooperation level. In other words, if we want the whole population to require and waste fewer goods, we may allow them to use more individually when just a few will do it properly. But this result, of course, only holds for small $\lambda$ values when the social dilemma is strong. The opposite behavior is true for a large $\lambda$ (weak dilemma), which meets our expectations.

As we noted above, $m$ plays the same role as selection strength in the Fermi-function, hence by increasing $m$ we increase the selection strength. This is shown on the top of all panels in Fig.~\ref{graph c}. This interpretation suggests that the cooperation level changes just until a certain level of selection, above this, for higher $m$, to strengthen the selection further has no particular consequence on the cooperation level.

\section{Evolutionary dynamics for heterogeneous populations}\label{D}

\begin{figure}[h!]
	\centering
	\includegraphics[width=14.5cm]{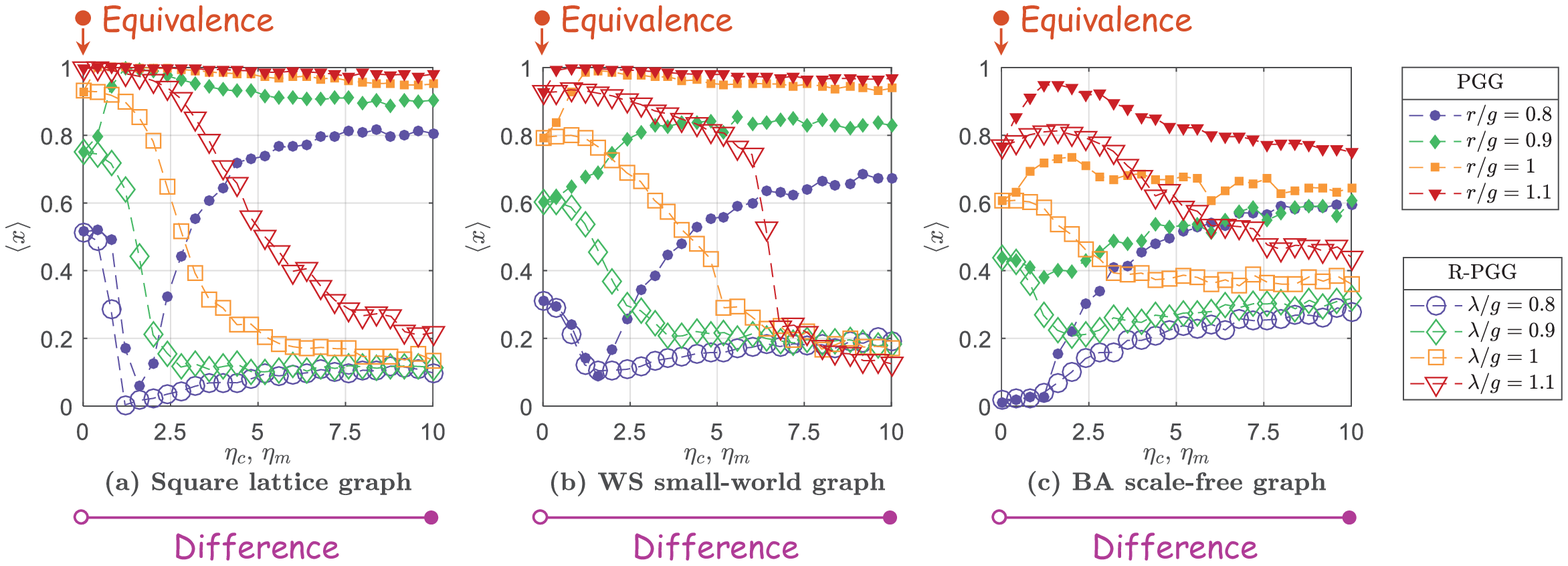}\\
	\caption{The $\langle x\rangle$ cooperation level in dependence on the contribution's heterogeneity $\eta_c$ (for PGG) or the requirement's heterogeneity $\eta_{m}$ (for R-PGG). Panels show the results for different topologies. W can observe conceptually similar differences between the two models as we presented in Fig.~\ref{graph hetero r}. We fix $c=m=1$.}\label{graph hetero c}
\end{figure}

We here present our numerical results obtained for heterogeneous populations. First, Fig.~\ref{graph hetero c} shows the cases when heterogeneity is introduced via individual cost or requirement values. Accordingly, the key parameter is $\eta_c$ for PGG or $\eta_m$ for R-PGG models. The system behaviors, hence the conclusions, are more or less similar to those we obtained when alternative heterogeneity was used via $\eta_r$ or $\eta_\lambda$ in Sec.~\ref{hetestruc}. As we already stressed, when $\eta_c=\eta_m=0$, means in the homogeneous case, R-PGG and PGG are equivalent. But their difference becomes striking as we increase the heterogeneity. More precisely, if the dilemma is hard, at low $r$ and low $\lambda$ values, then there is a small initial interval in heterogeneity where both PGG and R-PGG behave similarly. In this limited parameter region the cooperation level decreases by enlarging the heterogeneity among players. But beyond a threshold value the difference between the models starts growing again. Comparing with the heterogeneous $\eta_r$ and $\eta_\lambda$ cases, discussed in the main text, we here also observe a convergence toward the same cooperation level independently of the actual value of $r$ or $\lambda$ which characterizes the dilemma strength. But the convergence is significantly faster for R-PGG here, especially for square lattice topology.

\begin{figure}[h!]
	\centering
	\includegraphics[width=13.5cm]{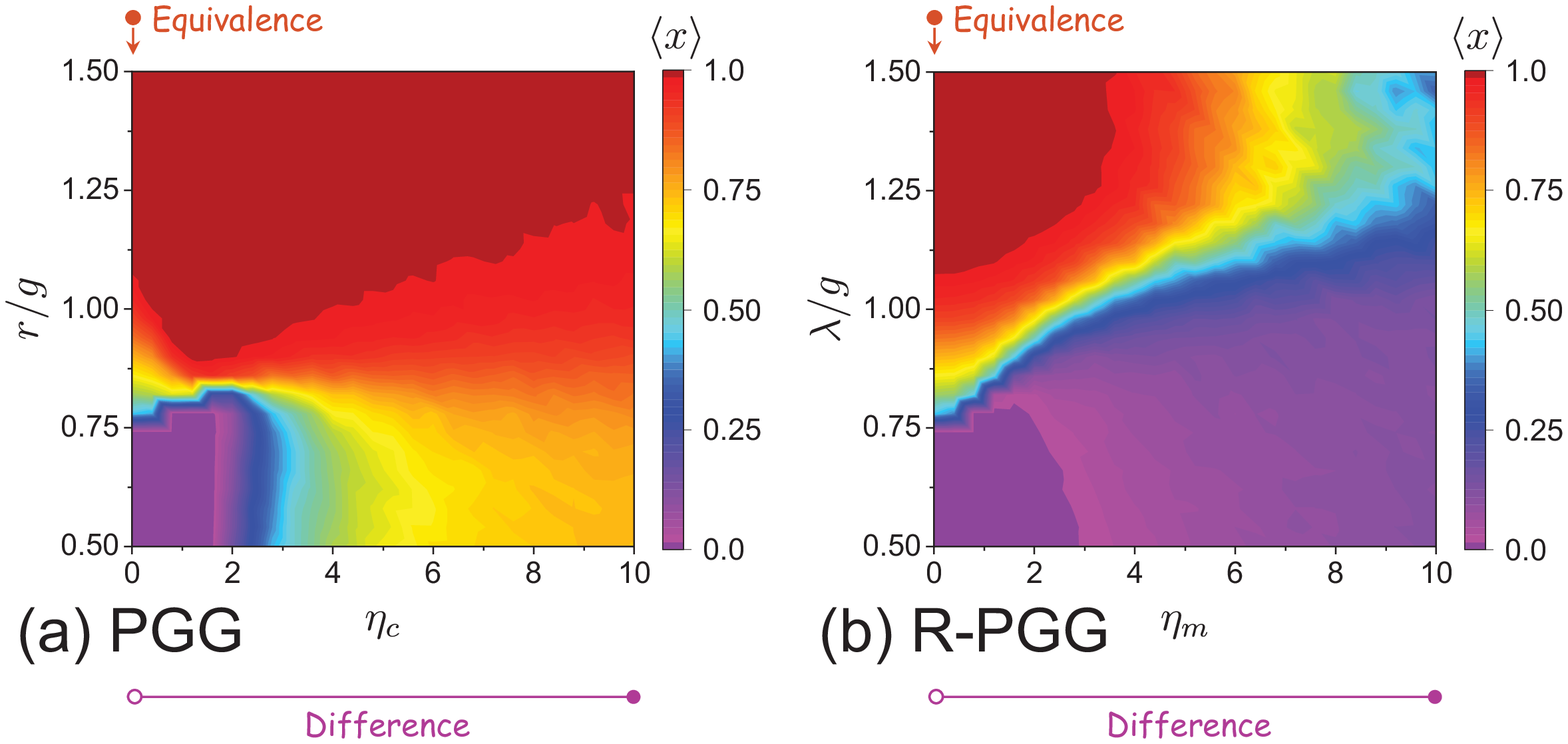}\\
	\caption{
		(a) The heat map of cooperation level on the parameter plane of normalized synergy factor $r/g$ and the heterogeneous $\eta_c$ contribution in PGG. (b) The same plot for R-PGG on the parameter plane of normalized waste factor $\lambda/g$ and requirement's heterogeneity $\eta_m$. The simulations are performed on the square lattice graph. While heterogeneity may have a positive consequence for PGG in the low $r$ region, this effect is completely missing in panel~(b) for R-PGG.  We fix $c=m=1$.}\label{graph r hetero c}
\end{figure}

As we have done previously, the whole range of $r$ and $\lambda$ can be studied by using the same range of heterogeneity. This can be seen in Fig.~\ref{graph r hetero c} where we present the general cooperation level as a heat map on $r-\eta_c$ and $\lambda-\eta_m$ parameter planes. The comparison of these panels supports our original observation. In particular, while heterogeneity can be useful for PGG to elevate the cooperation level, it is always detrimental in R-PGG. And this negative effect can be observed for any $\lambda$ value.
\begin{figure}[h!]
	\centering
	\includegraphics[width=13cm]{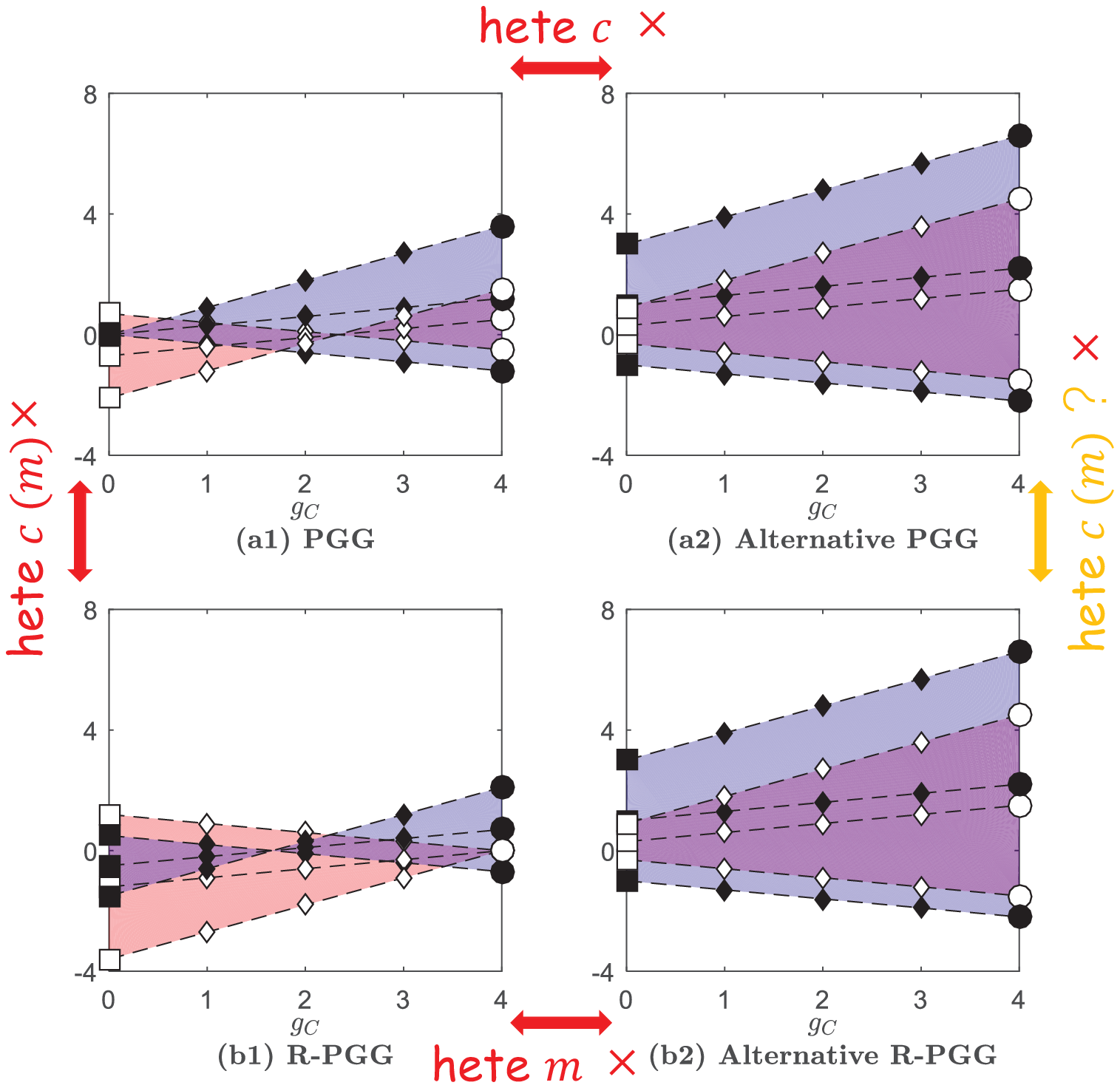}\\
	\caption{Payoff values for a cooperator (open symbols) and for a defector (closed symbols) player in dependence on the number of other cooperator players in the group. Panels show cases specified in Fig.~\ref{figureasymmetric} and in Fig.~\ref{figureasymmetric heter}. While $g=5$, $r=\lambda=1.5$ parameters are fixed, the three curves for each $\pi_C$ or $\pi_D$ are plotted by taking $c_i=m_i=1-2$, $1$, $1+2$ ($c=m=1$, $\eta_c=\eta_m=2$). The shaded areas mark the traces of curves when $c_i$ or $m_i$ values changes in the mentioned interval bordered by $c\pm\eta_c$ ($m\pm\eta_{m}$). These areas are marked by different colors for competing strategies (purple for $D$ and pink for $C$). The meaning of color arrows are similar to those we used in Fig.~\ref{figureasymmetric} in the main text.}\label{figureasymmetric hetec}
\end{figure}

\begin{figure}[h!]
	\centering
	\includegraphics[width=13.5cm]{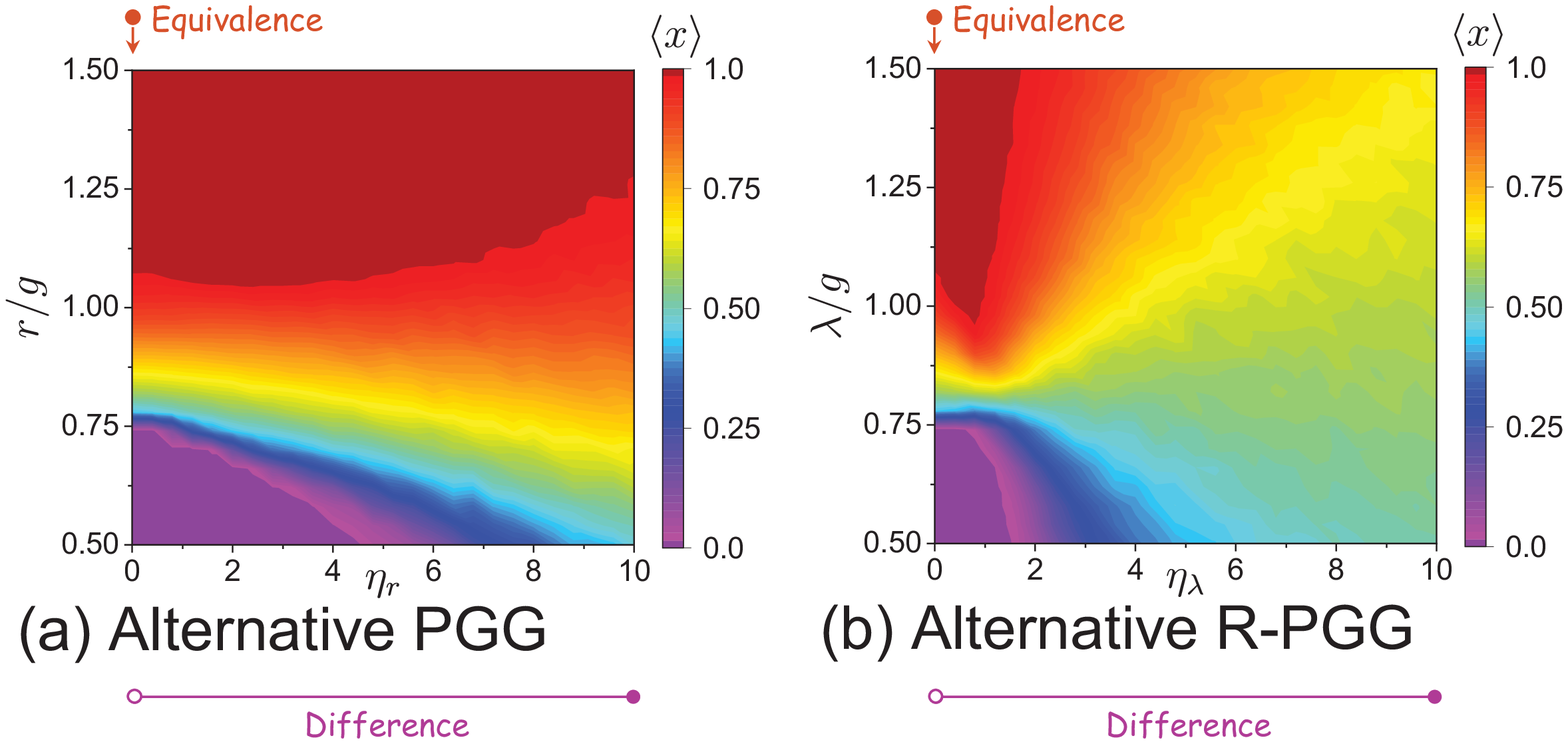}\\
	\caption{
		(a) The heat map of cooperation level on the parameter plane of average synergy factor $r/g$ and the amplitude of $\eta_r$ in alternative PGG. (b) Similar plot for alternative R-PGG on the parameter plane of average waste factor and the amplitude of  $\eta_{\lambda}$. Note that the results of alternative PGG (panel~(a)) are equivalent to PGG (Fig.~\ref{graph r hetero r}~(a)). Intriguingly, the results of alternative PGG (panel~(a)) are different from alternative R-PGG (panel~(b)). The simulations are performed on the square lattice graph. We fix $c=m=1$.}\label{figurealter r}
\end{figure}

\begin{figure}[h!]
	\centering
	\includegraphics[width=13.5cm]{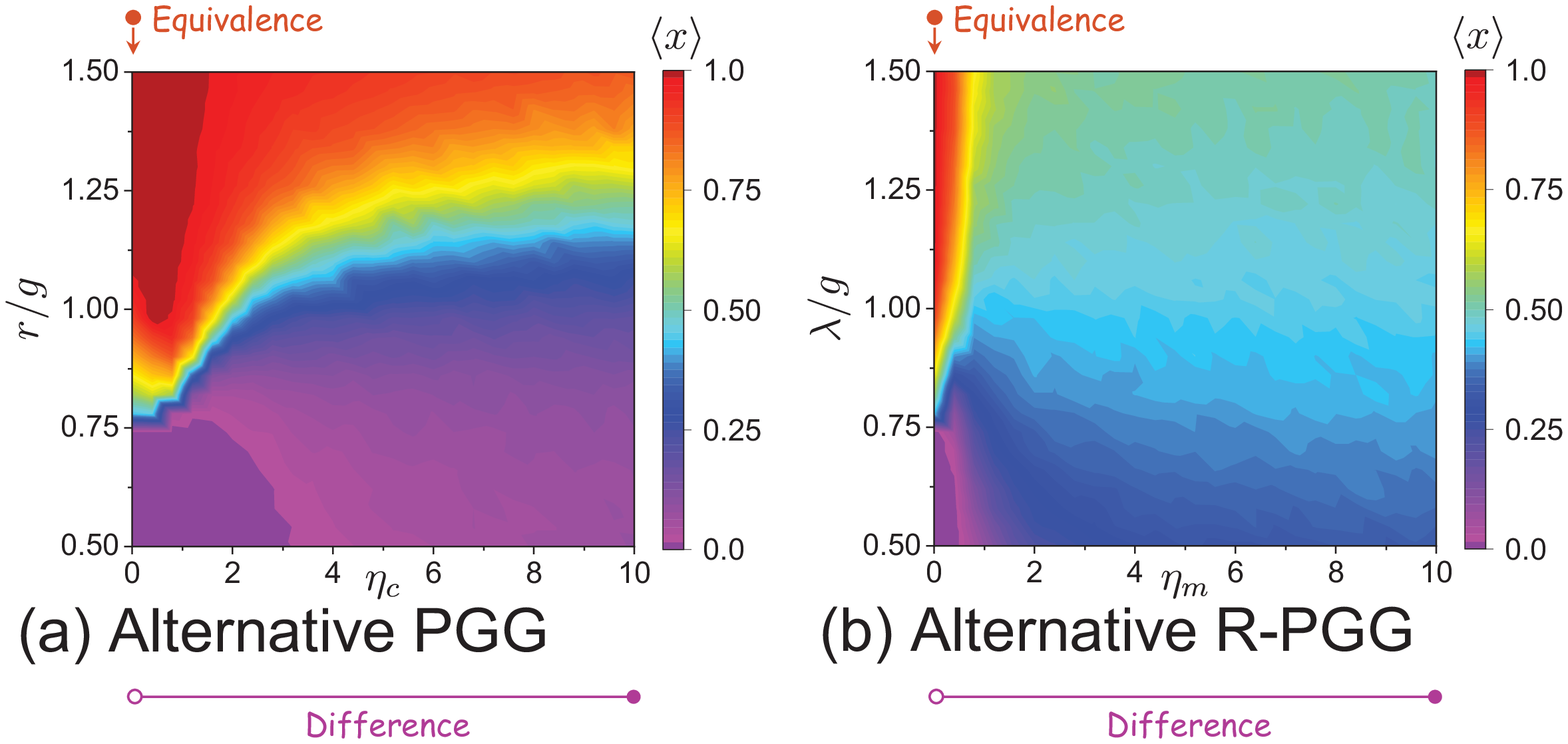}\\
	\caption{
		(a) The heat map of cooperation level on the parameter plane of normalized synergy factor $r/g$ and the amplitude of $\eta_c$ in alternative PGG. (b) The same plot for alternative R-PGG on the plane of normalized waste factor $\lambda/g$ and the amplitude of requirement's heterogeneity $\eta_m$. Interestingly, the results of alternative PGG (panel~(a)) are different from alternative R-PGG (panel~(b)). The simulations are performed on the square lattice graph. We fix $c=m=1$.}\label{figurealter c}
\end{figure}

To complete our comparison, in Fig.~\ref{figureasymmetric hetec} we present the consequence of heterogeneity in $c$ or $m$ parameter on the payoff functions. Similarly to the previously discussed situation, there are conceptual differences among the four cases. As panel Fig.~\ref{figureasymmetric hetec}(a1) and Fig.~\ref{figureasymmetric hetec}(a2) show, PGG and alternative PGG are different. This is because their coefficients in front of $c$ are distinct: in $\pi_C$, the coefficient is $((g_C+1)r/g-1)$ for PGG and $(g_C+1)r/g$ for alternative PGG; in $\pi_D$, the coefficient is $g_C r/g$ for PGG and $(g_C r/g+1)$ for alternative PGG. As a result, the same level of change in individual $c$ leads to a varying change in their payoff functions, resulting in different shapes of shaded areas. Nevertheless, both shapes and the absolute value of payoff functions for alternative PGG (shown in Fig.~\ref{figureasymmetric hetec}(a2)) and for alternative R-PGG (depicted in Fig.~\ref{figureasymmetric hetec}(b2)) are still the same. However, the resulting cooperation levels are different, as shown in Fig.~\ref{figurealter c}.
We hence stress that monitoring how payoff functions vary is not enough to reveal all differences among the cases we discussed above. Therefore, to complete the results presented in Fig.~\ref{graph r hetero r} and in Fig.~\ref{graph r hetero c}, we also measure the general cooperation level for alternative PGG and alternative R-PGG. These results are summarized in Fig.~\ref{figurealter r} and in Fig.~\ref{figurealter c}. The comparison of Fig.~\ref{figurealter r}(a) and Fig.~\ref{graph r hetero r}(a) confirms that the original PGG and the alternative PGG behave similarly when individual $r_i$ is introduced via nonzero $\eta_r$. 

Similarly, Fig.~\ref{figurealter c}(a) and Fig.~\ref{graph r hetero c}(a) show that PGG and alternative PGG are different when heterogeneous $c$ values are introduced via nonzero $\eta_c$. 
Importantly, the comparison of Fig.~\ref{figurealter r}(a) and Fig.~\ref{figurealter r}(b) highlights that the cooperation levels in heterogeneous alternative PGG and in alternative R-PGG are different. No matter whether their payoff values vary similarly by changing the average $r$, $\lambda$ values. Hence, we can confirm that alone the payoff values and how they change by varying key parameters, i.e. the shape of the shaded area, cannot provide sufficient information to compare different games.

\section*{References}


\begin{thebibliography}{10}
	\expandafter\ifx\csname url\endcsname\relax
	\def\url#1{{\tt #1}}\fi
	\expandafter\ifx\csname urlprefix\endcsname\relax\def\urlprefix{URL }\fi
	\providecommand{\eprint}[2][]{\url{#2}}

\bibitem{milinski_pnas08}
Milinski M, Sommerfeld R~D, Krambeck H~J, Reed F~A and Marotzke J 2008 {\it
	Proc. Natl. Acad. Sci. USA\/} {\bf 105} 2291--2294

\bibitem{hilbe_prsb10}
Hilbe C and Sigmund K 2010 {\it Proc. R. Soc. B\/} {\bf 277} 2427--2433

\bibitem{pacheco_plrev14}
Pacheco J~M, Vasconcelos V~V and Santos F~C 2014 {\it Physics of Life
	Reviews\/} {\bf 11} 573--586

\bibitem{sun_ww_is21}
Sun W, Liu L, Chen X, Szolnoki A and Vasconcelos V~V 2021 {\it iScience\/} {\bf
	24} 102844

\bibitem{szabo1998evolutionary}
Szab{\'o} G and T{\H{o}}ke C 1998 {\it Phys. Rev. E\/} {\bf 58} 69

\bibitem{takeshue_epl19}
Takeshue H 2019 {\it EPL\/} {\bf 126} 58001

\bibitem{amaral_pre21}
Amaral M~A and de~Oliveira M~M 2021 {\it Phys. Rev. E\/} {\bf 104} 064102

\bibitem{zhu_pc_epjb21}
Zhu P, Hou X, Guo Y, Xu J and Liu J 2021 {\it Eur. Phys. J. B\/} {\bf 94} 58

\bibitem{hauert2004spatial}
Hauert C and Doebeli M 2004 {\it Nature\/} {\bf 428} 643--646

\bibitem{chen_xj_epl10}
Chen X and Wang L 2010 {\it EPL\/} {\bf 90} 38003

\bibitem{graeser_njp11}
Gr{\"a}ser O, Xu C and Hui P~M 2011 {\it EPL\/} {\bf 13} 083015

\bibitem{skyrms2004stag}
Skyrms B 2004 {\it Stag-Hunt Game and the Evolution of Social Structure\/}
(Cambridge, U.K.: Cambridge Univ. Press)

\bibitem{starnini_jsm11}
Starnini M, S{\'a}nchez A, Poncela J and Moreno Y 2011 {\it J. Stat. Mech.\/}
{\bf 2011} P05008

\bibitem{deng_ys_epjb22}
Deng Y and Zhang J 2022 {\it Eur. Phys. J. B\/} {\bf 95} 29

\bibitem{page2000spatial}
Page K~M, Nowak M~A and Sigmund K 2000 {\it Proc. R. Soc. B\/} {\bf 267}
2177--2182

\bibitem{sinatra_jstat09}
Sinatra R, Iranzo J, G{\'o}mez-Garde{\~n}es J, Flor\'{\i}a L~M, Latora V and
Moreno Y 2009 {\it J. Stat. Mech.\/} {\bf 2009} P09012

\bibitem{szolnoki_prl12}
Szolnoki A, Perc M and Szab{\'o} G 2012 {\it Phys. Rev. Lett.\/} {\bf 109}
078701

\bibitem{chen_w_pa19}
Chen W, Wu T, Li Z and Wang L 2019 {\it Physica A\/} {\bf 519} 319--325

\bibitem{berg1995trust}
Berg J, Dickhaut J and McCabe K 1995 {\it Games and Econ. Behav.\/} {\bf 10}
122--142

\bibitem{chica_srep19}
Chica M, Chiong R, Adam M~T~P and Teubner T 2019 {\it Sci. Rep.\/} {\bf 9}
19789

\bibitem{zheng_lp_pa21}
Zheng L, Xu H, Tian C and Fan S 2021 {\it Physica A\/} {\bf 581} 126228

\bibitem{nowak2006evolutionary}
Nowak M~A 2006 {\it Evolutionary dynamics: exploring the equations of life\/}
(Harvard University Press)

\bibitem{li_xy_epjb20}
Li X, Chen T, Chen Q and Zhang X 2020 {\it Eur. Phys. J. B\/} {\bf 93} 204

\bibitem{yang_hx_pa20}
Yang H~X and Sun L 2020 {\it Physica A\/} {\bf 540} 123255

\bibitem{wang2021replicator}
Wang C and Hui K 2021 {\it Phys. Lett. A\/} {\bf 420} 127759

\bibitem{wang2022modeling}
Wang C, Huang C, Pan Q and He M 2022 {\it Chaos, Solit. \& Fract.\/} {\bf 158}
112092

\bibitem{wang_cq_amc22}
Wang C and Szolnoki A 2022 {\it Appl. Math. Comput.\/} {\bf 430} 127307

\bibitem{zheng2007cooperative}
Zheng D~F, Yin H, Chan C~H and Hui P 2007 {\it EPL\/} {\bf 80} 18002

\bibitem{souza2009evolution}
Souza M~O, Pacheco J~M and Santos F~C 2009 {\it J. Theor. Biol.\/} {\bf 260}
581--588

\bibitem{li_k_csf21}
Li K, Mao Y, Wei Z and Cong R 2021 {\it Chaos, Solit. and Fract.\/} {\bf 143}
110591

\bibitem{pacheco2009evolutionary}
Pacheco J~M, Santos F~C, Souza M~O and Skyrms B 2009 {\it Proc. R. Soc. B\/}
{\bf 276} 315--321

\bibitem{luo2021evolutionary}
Luo Q, Liu L and Chen X 2021 {\it Physica D\/} {\bf 424} 132943

\bibitem{chen_w_srep17}
Chen W, Gracia-L{\'a}zaro C, Li Z, Wang L and Moreno Y 2017 {\it Sci. Rep.\/}
{\bf 7} 4800

\bibitem{chica_cnsns19}
Chica M, Chiong R, Ramasco J~J and Abbass H 2019 {\it Commun. Nonlin. Sci.
	Numer. Sim.\/} {\bf 79} 104870

\bibitem{sigmund2010calculus}
Sigmund K 2010 {\it The calculus of selfishness\/} (Princeton University Press)

\bibitem{hilbe_n18}
Hilbe C, \v{S} \v{S}imca, Chatterjee K and Nowak M~A 2018 {\it Nature\/} {\bf
	559} 246--249

\bibitem{couto_njp22}
Couto M~C, Giaimo S and Hilbe C 2022 {\it New J. Phys.\/} {\bf 24} 063010

\bibitem{perc2013evolutionary}
Perc M, G{\'o}mez-Gardenes J, Szolnoki A, Flor{\'\i}a L~M and Moreno Y 2013
{\it J. R. Soc. Interface\/} {\bf 10} 20120997

\bibitem{alvarez2021evolutionary}
Alvarez-Rodriguez U, Battiston F, de~Arruda G~F, Moreno Y, Perc M and Latora V
2021 {\it Nat. Human Behav.\/} {\bf 5} 586--595

\bibitem{lv_amc22}
Lv S and Song F 2022 {\it Appl. Math. Comput.\/} {\bf 412} 126586

\bibitem{kang_hw_pla21}
Kang H, Zhou X, Shen Y, Sun X and Chen Q 2021 {\it Phys. Lett. A\/} {\bf 417}
127678

\bibitem{shen_y_pla22}
Shen Y, Yin W, Kang H, Zhang H and Wang M 2022 {\it Phys. Lett. A\/} {\bf 428}
127935

\bibitem{liu_lj_rspa22}
Liu L and Chen X 2022 {\it Proc. R. Soc. A\/} {\bf 478} 20220290

\bibitem{gaechter17}
G{\"a}chter S, K{\"o}lle F and Quercia S 2017 {\it Nat. Hum. Behav.\/} {\bf 1} 650--656

\bibitem{nowak2004emergence}
Nowak M~A, Sasaki A, Taylor C and Fudenberg D 2004 {\it Nature\/} {\bf 428}
646--650

\bibitem{quan_j_pa21}
Quan J, Tang C and Wang X 2021 {\it Physica A\/} {\bf 563} 125488

\bibitem{su2018understanding}
Su Q, Wang L and Stanley H~E 2018 {\it New J. Phys.\/} {\bf 20} 103030

\bibitem{fu_mj_pa21}
Fu M, Wang J, Cheng L and Chen L 2021 {\it Physica A\/} {\bf 580} 125672

\bibitem{ohdaira_srep22}
Ohdaira T 2022 {\it Sci. Rep.\/} {\bf 12} 6604

\bibitem{schuster1983replicator}
Schuster P and Sigmund K 1983 {\it J. Theor. Biol.\/} {\bf 100} 533--538

\bibitem{duong_dga20}
Duong M~H and Han T~A 2020 {\it Dyn. Games Appl.\/} {\bf 10} 641--663

\bibitem{wang_q_amc18}
Wang Q, He N and Chen X 2018 {\it Appl. Math. Comput.\/} {\bf 328} 162--170

\bibitem{liang_rh_pre22}
Liang R, Wang Q, Zhang J, Zheng G, Ma L and Chen L 2022 {\it Phys. Rev. E\/}
{\bf 105} 054302

\bibitem{nowak1992evolutionary}
Nowak M~A and May R~M 1992 {\it Nature\/} {\bf 359} 826--829

\bibitem{ohtsuki2006simple}
Ohtsuki H, Hauert C, Lieberman E and Nowak M~A 2006 {\it Nature\/} {\bf 441}
502--505

\bibitem{perc2017statistical}
Perc M, Jordan J~J, Rand D~G, Wang Z, Boccaletti S and Szolnoki A 2017 {\it
	Phys. Rep.\/} {\bf 687} 1--51

\bibitem{allen2017evolutionary}
Allen B, Lippner G, Chen Y~T, Fotouhi B, Momeni N, Yau S~T and Nowak M~A 2017
{\it Nature\/} {\bf 544} 227--230

\bibitem{flores_jtb21}
Flores L~S, Fernandes H~C, Amaral M~A and Vainstein M~H 2021 {\it J. Theor.
	Biol.\/} {\bf 524} 110737

\bibitem{quan_j_c19}
Quan J, Li X and Wang X 2019 {\it Chaos\/} {\bf 29} 103137

\bibitem{wei_x_epjb21}
Wei X, Xu P, Du S, Yan G and Pei H 2021 {\it Eur. Phys. J. B\/} {\bf 94} 210

\bibitem{perc_pre08}
Perc M and Szolnoki A 2008 {\it Phys. Rev. E\/} {\bf 77} 011904

\bibitem{santos_n08}
Santos F~C, Santos M~D and Pacheco J~M 2008 {\it Nature\/} {\bf 454} 213--216

\bibitem{lei2010heterogeneity}
Lei C, Wu T, Jia J~Y, Cong R and Wang L 2010 {\it Physica A\/} {\bf 389}
4708--4714

\bibitem{szolnoki_amc20}
Szolnoki A and Chen X 2020 {\it Appl. Math. Comput.\/} {\bf 385} 125430

\bibitem{lee_hw_pa21}
Lee H~W, Cleveland C and Szolnoki A 2021 {\it Physica A\/} {\bf 582} 126222

\bibitem{liu_jz_epjb21}
Liu J, Peng M, Peng Y, Li Y, Chu C, Li X and Liu Q 2021 {\it Eur. Phys. J. B\/}
{\bf 94} 167

\bibitem{rong_zh_c19}
Rong Z, Wu Z~X, Li X, Holme P and Chen G 2019 {\it Chaos\/} {\bf 29} 103103

\bibitem{deng_ys_pa21}
Deng Y and Zhang J 2021 {\it Physica A\/} {\bf 584} 126363

\bibitem{yuan2014role}
Yuan W~J and Xia C~Y 2014 {\it PLoS ONE\/} {\bf 9} e91012

\bibitem{huang2015effect}
Huang K, Wang T, Cheng Y and Zheng X 2015 {\it PLoS ONE\/} {\bf 10} e0120317

\bibitem{weng2021heterogeneous}
Weng Q, He N, Hu L and Chen X 2021 {\it Phys. Lett. A\/} {\bf 400} 127299

\bibitem{ma2021effect}
Ma X, Quan J and Wang X 2021 {\it Chaos, Solit. \& Fract.\/} {\bf 152} 111353

\bibitem{zhang2017impact}
Zhang Y, Wang J, Ding C and Xia C 2017 {\it Knowledge-Based Syst.\/} {\bf 136}
150--158

\bibitem{fan2017promotion}
Fan R, Zhang Y, Luo M and Zhang H 2017 {\it Physica A\/} {\bf 465} 454--463

\bibitem{liu2021effects}
Liu J, Peng M, Peng Y, Li Y, Chu C, Li X and Liu Q 2021 {\it Eur. Phys. J. B\/}
{\bf 94} 1--7

\bibitem{hauser2019social}
Hauser O~P, Hilbe C, Chatterjee K and Nowak M~A 2019 {\it Nature\/} {\bf 572}
524--527

\bibitem{mcavoy2020social}
McAvoy A, Allen B and Nowak M~A 2020 {\it Nat. Human Behav.\/} {\bf 4} 819--831

\bibitem{su2019spatial}
Su Q, Li A, Wang L and Eugene~Stanley H 2019 {\it Proc. R. Soc. B\/} {\bf 286}
20190041

\bibitem{szolnoki_pre09c}
Szolnoki A, Perc M and Szab{\'o} G 2009 {\it Phys. Rev. E\/} {\bf 80} 056109

\bibitem{gracia-lazaro_pre14}
Gracia-L{\'a}zaro C, G{\'o}mez-Garde{\~n}es J, Flor{\' \i}a L~M and Moreno Y
2014 {\it Phys. Rev. E\/} {\bf 90} 042808

\bibitem{yang_hx_pa19}
Yang H~X and Yang J 2019 {\it Physica A\/} {\bf 523} 886--893

\bibitem{meloni_rsos17}
Meloni S, Xia C~Y and Moreno Y 2017 {\it R. Soc. open sci.\/} {\bf 4} 170092

\bibitem{szolnoki_epl10}
Szolnoki A and Perc M 2010 {\it EPL\/} {\bf 92} 38003

\bibitem{watts1998collective}
Watts D~J and Strogatz S~H 1998 {\it Nature\/} {\bf 393} 440--442

\bibitem{barabasi1999emergence}
Barab{\'a}si A~L and Albert R 1999 {\it Science\/} {\bf 286} 509--512

\bibitem{karlin2014first}
Karlin S 2014 {\it A first course in stochastic processes\/} (Academic Press)

\bibitem{allen2014games}
Allen B and Nowak M~A 2014 {\it EMS Surv. Math. Sci.\/} {\bf 1} 113--151

\end{thebibliography}
\providecommand{\newblock}{}

\end{document}